\documentclass[fleqn,usenatbib]{mnras}
\usepackage[T1]{fontenc}
\usepackage{ae,aecompl}

\usepackage{graphicx}	% Including figure files
\usepackage{amsmath}	% Advanced maths commands
\usepackage{amssymb}	% Extra maths symbols

\usepackage{amsmath}
\usepackage{float}
\usepackage{tikz}
\usepackage{listings}
\usepackage{caption}
\usepackage{animate}
\usepackage{subcaption}
\usetikzlibrary{positioning}
\usepackage[utf8]{inputenc}
\usepackage{setspace}
\usepackage{bm}
\usepackage{caption}
\usepackage{color}
\usepackage{soul}

\newcommand*    \msun{{\,\mathrm{M}_{\odot}}}
\newcommand*    \kms{{\, \rm km\,s^{-1}}}
\newcommand*    \pc{{\, \mathrm{pc}}}
\newcommand*    \kpc{{\, \mathrm{kpc}}}
\newcommand*    \myr{{\, \rm Myr}}
\newcommand*    \msbh{M_\bullet}
\newcommand*    \mdm{M_{\mathrm{DM}}}
\newcommand*    \ndm{N_{\mathrm{DM}}}
\newcommand*    \rmo{r_{\mathrm{m}}}
\newcommand*    \af{a_{\mathrm{f}}}
\newcommand*    \ah{a_{\mathrm{h}}}

\title[Carving the largest observed galactic cores]{Formation of the largest galactic cores through binary scouring and gravitational wave recoil}

\author[I. T. Nasim et al.]{%
Imran Nasim,$^{1}$\thanks{E-mail: i.nasim@surrey.ac.uk (KTS)}
Alessia Gualandris,$^{\!1}$
Justin I. Read,$^{\!1}$
Fabio Antonini,$^{\!1,2}$
\newauthor
Walter Dehnen,$^{\!3,4,5}$ 
and Maxime Delorme$^{1,6}$
\\
$^{1}$ Department of Physics, University of Surrey, Guildford, GU2 7XH, Surrey, UK\\
$^{2}$ School of Physics and Astronomy, Cardiff University, Cardiff, CF24 3AA, UK\\
$^{3}$ Astronomisches Rechen-Institut, Zentrum für Astronomie der Universität Heidelberg, Mönchhofstr.~12-14, 69120, Heidelberg, Germany\\
$^{4}$ Universit{\"a}ts-Sternwarte M\"unchen, Scheinerstr.~1, 81679, Munich, Germany\\
$^{5}$ School of Phyiscs and Astronomy, University of Leicester, University Rd, LE1 7RH, UK\\
$^{6}$ D\'epartement d'Astrophysique/AIM, CEA/IRFU, CNRS/INSU, Universit\'e Paris-Saclay, Universit\'e  de Paris, 91191 Gif-sur-Yvette, France
}

\date{}
\pubyear{2020}

\begin{document}
\label{firstpage}
\pagerange{\pageref{firstpage}--\pageref{lastpage}}
\maketitle

\begin{abstract}
Massive elliptical galaxies are typically observed to have central cores in their projected radial light profiles. Such cores have long been thought to form through `binary scouring' as supermassive black holes (SMBHs), brought in through mergers, form a hard binary and eject stars from the galactic centre. However, the most massive cores, like the $\sim 3\kpc$ core in A2261-BCG, remain challenging to explain in this way. In this paper, we run a suite of dry galaxy merger simulations to explore three different scenarios for central core formation in massive elliptical galaxies: `binary scouring', `tidal deposition' and `gravitational wave (GW) induced recoil'. Using the \textsc{griffin} code, we  self-consistently model the stars, dark matter and SMBHs in our merging galaxies, following the SMBH dynamics through to the formation of a hard binary. We find that we can only explain the large surface brightness core of A2261-BCG with a combination of a major merger that produces a small $\sim 1\kpc$ core through binary scouring, followed by the subsequent GW recoil of its SMBH that acts to grow the core size. Key predictions of this scenario are an offset SMBH surrounded by a compact cluster of bound stars and a non-divergent central density profile.  We show that the bright `knots' observed in the core region of A2261-BCG are best explained as stalled perturbers resulting from minor mergers, though the brightest may also represent ejected SMBHs surrounded by a stellar cloak of bound stars. 

\end{abstract}


%%%%%%%%%%%%%%%%% BODY OF PAPER %%%%%%%%%%%%%%%%%%

\section{Introduction}
\label{sec:intro}
Observations by the \textit{Hubble Space Telescope} (HST) have revealed a clear bimodality in the population of elliptical galaxies, with brighter ellipticals ($M_B\leq -21$) tending to have ``cored" central surface brightness profiles and fainter ellipticals ($M_B>-21$) showing ``cuspy" central profiles instead \citep{kingminkowski1966,faber1997,graham2019}. 
While cored galaxies tend to have a small degree of rotation with boxlike isophotes, cuspy galaxies have disky isophotes with a faster rotation \citep{kormendybender1996,lauer2007}.
\citet{krajnovic2013} show that the majority of massive, slowly rotating galaxies have surface brightness cores and that all slow rotating galaxies with stellar masses above $2 \times {10^{11}}\msun$ have such cores. This suggests that the observed structural differences arise due to different formation pathways of elliptical galaxies.\newline
In the galactic merger scheme presented by \citet{NaabOstriker2017}, massive elliptical galaxies, which are generally gas-poor, form through a multi stage process where the early evolution  is dominated by in-situ star formation accompanied by multiple accretion events with star-bursting progenitors. Subsequent growth ($z\leq 3$) is dominated by dry (gas-poor) mergers, which has been shown to rapidly grow the size of ellipticals \citep{Naab2009}. Fainter ellipticals, on the other hand, form through in-situ star formation along with multiple accretion events with gas-rich disk galaxies leading to steeply rising central surface brightness profiles \citep{NaabTrujillo2006,JohanssonNaab2009}.
While this framework explains many observed properties, the existence of surface brightness cores in the most massive ellipticals poses a challenging theoretical problem. If massive ellipticals form from the merger of fainter cuspy galaxies \citep{feldmann2011,Moster2013}, they should exhibit cuspy profiles, as it has been shown that mergers preserve the steep density cusps of the merging galaxies \citep[e.g.][]{boychainkolchin2004,dehnen05}.\newline
Currently the most accepted mechanism for core formation in ellipticals is scouring 
by binary supermassive black holes (SMBHs) \citep{begelman1980,hills1983,quinlan1996}. During the merger of two gas-poor galaxies, the central SMBHs sink to the centre of the remnant due to dynamical friction \citep{chandrasekhar1943} until they form a gravitationally bound pair.
The massive black hole binary (BHB) interacts with the surrounding stellar environment, causing the ejection of stars on intersecting orbits. This removes energy and angular momentum from the binary, causing it to harden \citep{hills1983,quinlan1996,sesana2006}, and carves a core in the stellar distribution.
If the binary shrinks to a separation, typically of the order of milliparsecs, where emission of gravitational waves (GWs) becomes important, the evolution proceeds quickly to inspiral and coalescence to a single SMBH. Anisotropic emission of GWs imparts a kick velocity, known as GW recoil, to the newly formed black hole.
Thanks to the recent advancements in numerical relativity \citep{Campanelli2006,lousto2011,Healy2014,Healy2018}, it has become possible to simulate the merger of SMBHs in full general relativity and evaluate the size of GW kicks. Interestingly, recoil kicks range from a few hundred to $\sim 5000\kms$, depending on the configuration of the merging SMBHs and the relative orientation of their spins \citep{Campanelli2007,Gonzalez2007,lousto2011,lousto2019}. Large GW recoils remove the SMBHs from the central region, leading to damped oscillatory motion with repeated passages through the core. During such passages the SMBH injects energy into the central region, enlarging the core formed during the phase of binary hardening \citep[e.g.][]{boychainkolchin2004,gualandrismerritt2008,merritt2009}. The oscillations are slowly damped by dynamical friction until the SMBH reaches dynamical equilibrium with the background stellar distribution \citep{gualandrismerritt2008}. Because dynamical friction is inefficient in flat density profiles \citep[e.g.][]{read2006}, the oscillatory motion is long lived, and the core continues to be enlarged at each oscillation. While GW recoil can in principle lead to the ejection of the SMBH and its associated bound star cluster from the host galaxy \citep{merritt2009}, recoils exceeding the central escape speed (typically $\sim 1000$ $\rm{kms^{-1}}$ for massive galaxies \citep{devecchi2009}) are expected to be rare \citep{blecha2016}. A study by \citet{lousto2012} showed the probability of recoil velocities exceeding the escape velocity is $\sim 5\%$ for galaxies with escape velocities of $1000$ $\rm{kms^{-1}}$ and $\sim 20\%$ for galaxies with escape velocities of $500$ $\rm{kms^{-1}}$.\newline
Alternative processes for core formation include the ``Multiple SMBH scouring" mechanism by \citet{kulkarni2012},  a natural extension of the core-scouring mechanism to multiple SMBHs, the ``SMBH-AGN" scenario by \citet{martizzi2012} in which a core forms through repeated fluctuations of the inner gravitational potential on a timescale comparable to the local dynamical time\footnote{ Such a mechanism has been discussed extensively in the literature where the potential fluctuations owe to gas flows driven by stellar feedback \citep[e.g.][]{read2005,pontzen2012,read2019}. Here, gas flows are driven by AGN feedback instead. This scenario may not be able to produced large - kpc scale - cores but if possible such a large scale feedback will clear the inner galaxy of gas and hence prevent subsequent AGN activity.}, and the ``stalled perturber" scenario \citep{read2006,Goerdt2010}  where a sinking satellite transfers energy to the background galaxy causing a reduction in central density and the formation of a constant density core, with the satellite stalling at the edge of the core.
In practice, all of these mechanisms could act in tandem to a greater or lesser degree to drive the observed cores in massive ellipticals.\newline
Alongside the above theoretical uncertainties in the physics of core formation in ellipticals, there have also been observational challenges. In particular, determining the size of the core has proven to be a non trivial task.
The light profiles of ellipticals are well described by the 3-parameter S\'ersic profile \citep{sersic1963,sersic1968} over a large radial range. The most luminous ellipticals,
however, show a departure from the S\'ersic law in their central regions, at a radius widely known as the ``break" or ``core" radius. In these galaxies, the profiles break downward from the inward extrapolation of the outer S\'ersic law. 
Initially the core size of a galaxy was determined by fitting the so-called ``Nuker-profile" \citep{lauer1995} to the surface brightness profile, a method that however depends sensitively on the radial fitting range and yields unreliable results when fit to surface brightness profiles with a large radial extent \citep[e.g.][]{graham2003b,dullograham2012}. 
In more recent years, it has become customary to incorporate a central flattening in the light profile by adopting a 6-parameter core-S\'ersic profile \citep{graham2003b,trujillo2004} which provides a reliable measurement of the core size even over a large radial range \citep[e.g.][]{dullograham2012,dullograham2013,dullograham2014}. Furthermore, it has been shown that adopting a multi-component model rather than a single core-S\'ersic model over the entire radial range provides a more reliable estimate of the core size \citep{dullograham2014,dullo2019}.
Measured core sizes for massive ellipticals -- derived in this way -- are typically tens to a few hundred parsecs \citep[e.g.][]{dullograham2014}, while cores larger than $1\kpc$ are rare. A study by \citet{lauer2007} considered a large sample of BCGs and found that fewer than $10$ systems had a core size of $\sim 1\rm{kpc}$ or greater, with the largest cored system being NGC 6166 which has a core size of $\sim 1.5\kpc$. More recently, \citet{dullo2019} considered the largest sample of ``large-core" galaxies to date, finding that only 13(7) galaxies have core sizes larger than $0.5(1)\kpc$.\newline
Large-core galaxies are found to obey the same scaling relations between break radius and surface brightness or spheroid luminosity as normal core galaxies, as well as a log-linear break radius-SMBH mass relation, when a measurement is available \citep{dullo2019}. However, directly measured SMBHs are over-massive with respect to predictions from the $M_{\rm SMBH}-L$ and the $M_{\rm SMBH}-\sigma$ relations, with a more significant offset for the latter relation. This is generally interpreted as a result of an increased number of dry mergers for the most massive galaxies \citep[e.g.][]{dullo2019}.\newline
An extremely large surface brightness core has been found in the brightest cluster galaxy of Abell 2261 (A2261-BCG hereafter) \citep{postman2012}. Assuming a single component model, \citet{bonfini2016} infer a core-S\'ersic break radius of $3.6\kpc$ while assuming a multi-component model \citet{dullo2019} derive a core radius of $2.7\kpc$. With either estimate, this represents the second largest depleted core known to date \citep{dullo2019}.
The earlier measurement, in particular, is so extreme to imply an unrealistically large number of dry major mergers, at least in the context of the standard $\Lambda$CDM cosmology \citep{li2007,bonfini2016}.
Another peculiarity of A2261-BCG is that it contains a number of objects in the core region referred to as  ``knots", as they appear as high density regions in the surface brightness profile \citep{postman2012,bonfini2016}. \cite{2017ApJ...840...31D} suggested that these knots could be stalled satellites that decayed to the centre of the main galaxy through dynamical friction.
Together with the extremely large core size, the knots make this galaxy a unique system to study core formation scenarios.\newline
In this paper, we present a detailed numerical study of the effect of successive galactic dry mergers, both major and minor, on the core-size of ellipticals. Each merging galaxy hosts an SMBH and we model the mergers using the new \textsc{griffin} Fast Multiple Method \citep{dehnen02fmm,dehnen2014griffin}. This allows us to simultaneously capture the $N$-body dynamics of the merger and the SMBH-star and SMBH-SMBH interactions up to and after the formation of a hard binary \citep{nasim2020}. We select a sequence of dry mergers consistent with expectations in $\Lambda$CDM and representative of A2261-BCG. For a subset of our simulations, we additionally model the impact of the GW recoil of the newly formed SMBH following binary coalescence. Our key finding is that three distinct mechanisms are involved in the formation of the very largest surface brightness cores in ellipticals. The initial core is formed by binary scouring during the first dry major merger and is only minimally enlarged by subsequent major mergers. Gravitational wave recoil, on the other hand, can deplete the core further by ejecting additional stellar mass. Finally, minor mergers are responsible for the tidal deposition of material, including small satellites, on the edge the core. This further expands the size of the core and could plausibly explain the `knots' observed in A2261-BCG. We note that these cores are identified and measured by fitting core-S{\'e}rsic models to the surface density profiles, in analogy with the approach most commonly adopted by observers \citep[e.g.][]{graham2019}. The presence of a core, i.e. a region of (nearly) constant density, in the surface density profile of a galaxy does not imply a core in the three-dimensional density profile. Our models instead produce a shallow cusp in spatial density. \newline
This article is organised as follows. The numerical simulations and adopted galactic models are presented in Section \ref{numerical_sims}. The role of the first major merger is investigated in Section \ref{sec:firstmerger}, followed by a description of subsequent mergers combined with an analysis of core sizes and mass deficits in Section \ref{sec:multiple_mergers}. Gravitational wave recoil is studied in Section \ref{sec:GW_section}, where additional simulations of kicked SMBHs are presented. Our results on the relative importance and effects of the different core formation mechanisms are discussed in Section \ref{sec:discussion}. Finally, in Section \ref{sec:conclusion}, we present our conclusions on how we can best-explain the extremely large core observed in A2261-BCG based on the simulations presented here.

\section{Numerical Simulations}
\label{numerical_sims}
\subsection{Initial conditions}
\label{sec:initial} 
We consider multi-component galaxy models consisting of a stellar bulge, dark matter (DM) halo and a central massive black hole (BH). The bulge follows a S{\'e}rsic model \citep{sersic1963,sersic1968} in projection
\begin{equation} \label{EqSer}
I(R)=I_0 {\rm e}^{-b(R/R_{\rm e})^{1/n}}
\end{equation}
where $I_{0}$ is the central intensity, $R_{\rm e}$ is the (projected) effective half-light radius, $n$ describes the curvature of the outer profile and  $b$ is a function of $n$ given by $b = 2n - 1/3 + 0.009876/n$ \citep{terzicgraham2005}. The corresponding space density profile is  well approximated by \citep{terzicgraham2005} 
\begin{subequations}
\label{sersic_3d}
\begin{align}
&\rho (r) = \rho_0 \left(\frac{r} {R_{\rm e}}\right)^{-p}
{\rm e}^{-b\left( r/R_{\rm e} \right)^{1/n}}\\
&\rho_0 = \Upsilon  I_0 \ b^{n(1-p)}  \frac{\Gamma(2n)}{2 R_{\rm e} \Gamma(n(3-p))}  
\end{align}
\end{subequations}
\newline
where $p = 1.0 - 0.6097/n + 0.05563/n^2$ \citep{Marquez2000}, $\Gamma$ is the gamma function and for the mass to light ratio we assume $\Upsilon=3.5M_\odot/L_\odot$ \citep{bonfini2016}.

The dark matter halo follows a \cite*{nfw1996} profile (NFW)
\begin{equation}\label{nfwdens}
	\rho (r) = \frac{\rho_0}{(r/a)(1+r/a)^2},
\end{equation}
which is described by the two free parameters: $a$ and $\rho_{0}$. The scale length parameter $a$ is set equal to the scale radius $r_{\rm {s}} = r_{200}/c$, where $c$ is a concentration parameter and $r_{200}$ is the distance from the centre of the halo where the mean enclosed density is 200 times greater than the critical density of the Universe at redshift $z=0$, $\rho_{\rm {c}} = 136.05$\,M$_\odot$\,kpc$^{-3}$. We adopt the relation derived by \citet{dutton2014} to set the concentration parameter. The parameter $\rho_{0}$ is set by the mass enclosed within a sphere of radius the $r_{200}$, which is given by $M_{200}=200\rho_{c}\frac{4}{3}\pi r_{200}^3$.
The enclosed mass profile for Eq.~\eqref{nfwdens} is given by
\begin{equation}\label{nfw_mass}
M(r) = 4\pi \rho_{0}a^3\left[\ln{(1+r/a)} - \frac{r/a}{1+ r/a}\right].
\end{equation}
Because the NFW mass profile diverges logarithmically with radius we impose an outer-cutoff at $r_{200}$ with a logarithmic slope of $-5$.\newline
We use the \textsc{Agama} action-based modelling library \citep{agama2019}
to setup self consistent initial conditions for our
spherically symmetric,  multi-component, isotropic models. 
To this aim, we considered the combined potential of the stellar bulge, the dark matter halo and the central SMBH which is represented as a point mass $\msbh$ at rest at the origin. We then sampled each component from their distribution function to produce the multi-component models.

\subsection{Merger setup}
\label{ic_setup}
\begin{table}
\begin{center}
\caption{The core-S{\'e}rsic fit parameters (see Eq.\eqref{core_sersic}) for A2261-BCG from \citep{bonfini2016} (B16) and \citep{dullo2019} (D19).}
\label{tab:core_sersic_params}
\begin{tabular}{c c c c c c c} 
\hline 
Study & $\mu_{b}$ & $n$ & $\alpha$ & $\gamma$ & $R_{e}$ & $R_{b}$  \\
      & [mag $\rm{arcsec}^{-2}$]  &  &  &  & [kpc] & [kpc]   \\
\hline
B16 & 19.59 & 3.9  & 3.6  & 0.02 & 38.86 & 3.63 \\
D19 & 18.69 & 2.1  & 5.0  & 0.0  & 17.6 & 2.71 \\
\hline
\end{tabular} 
\end{center}
\end{table}
In order to investigate the origin of the largest observed cores we model mergers of massive elliptical galaxies including major and minor mergers and consider different sequences of such mergers.
We take A2261-BCG as our fiducial galaxy with an extremely large core, whose structural parameters were derived first by \citet{postman2012} and \citet{bonfini2016}, and more recently by \citet{dullo2019} by fitting a core-S{\'e}rsic model (see Table \ref{tab:core_sersic_params}).\newline
The initial conditions (ICs) of the progenitor galaxies are listed in Table \ref{tab:progenitor_params}. 
\begin{table*}
\begin{center}
\caption{Properties of the progenitor galaxies. From left to right, the columns give: stellar mass ($M_{\star}$), dark matter halo mass ($\mdm$), central SMBH mass ($\msbh$), effective radius ($R_{\rm{e}}$), S{\'e}rsic index ($n$), number of star particles ($N_{*}$) and number of DM particles ($\ndm$) in the realisations, brief progenitor description.}
\label{tab:progenitor_params}
\begin{tabular}{l c c c c c c c l} 
\hline 
\hline
Progenitor & $M_{\star}$ &  $\mdm$ & $\msbh$ & $R_{\rm {e}}$ & $n$ & $N_{*}$ & $\ndm$ & Description\\[0.8ex]
      & [$10^{10}\msun$]  & [$10^{10}\msun$]  & [$10^{10}\msun$] & [kpc] & & [$10^{6}$] & [$10^{6}$] &\\
\hline
IC-1  &  $2.2\times10^2$ &  $1.0\times10^4$ & $1.27$ &  $19.43$ & $3.9$ & $1.0$  & $1.0$  & 1:1 merger progenitor using B16 parameters  \\
IC-2  &  $2.2\times10^2$ &  $1.0\times10^4$ & $1.27$ &  $19.43$ & $3.9$ & $1.0$  & $10.0$ & Same as IC-1 with higher DM resolution  \\
IC-3  &  $2.2\times10^2$ &  $1.0\times10^4$ & $1.27$ &  $19.43$ & $3.9$ & $10.0$ & $1.0$  & Same as IC-1 with higher stellar resolution  \\
IC-4  & $1.6\times10^1$ &  $2.0\times10^3$ & $0.101$ &  $4.41$  & $4.0$ & $1.0$  & $1.0$  & 1:10 merger progenitor (abundance matching) \\
IC-5  &  $4.4\times10^1$ &  $2.0\times10^3$ & $0.254$ &  $8.42$  & $4.0$ & $1.0$  & $1.0$  & 1:10 merger progenitor (scaled from IC-1)  \\
IC-6  &  $1.1\times10^2$ &  $5.0\times10^3$ & $0.635$ &  $11.4$  & $4.0$ & $1.0$  & $1.0$  & 1:4 merger progenitor (scaled from IC-1)   \\
IC-7  &  $2.2\times10^2$ &  $1.0\times10^4$ & $1.27$ &  $8.8$   & $2.1$ & $1.0$  & $1.0$  & 1:1 merger progenitor using D19 parameters \\
IC-8  &  $2.2\times10^2$ &  $1.0\times10^4$ & $3.23$ &  $8.8$   & $2.1$ & $1.0$  & $1.0$  & Same as IC-7 but with larger SMBH mass  \\
\hline
IC-1a  &  $2.2\times10^2$ &  $1.0\times10^4$ & $0$ &  $19.43$ & $3.9$ & $1.0$  & $1.0$  & Same as IC-1 but with no SMBH \\
IC-1aE  &  $2.2\times10^2$ &  $1.0\times10^4$ & $0$ &  $19.43$ & $3.9$ & $1.0$  & $45.0$  & Same as IC-1 but with equal mass particles  \\ 
IC-6a  &  $1.1\times10^2$ &  $5.0\times10^3$ & $0$ &  $11.4$  & $4.0$ & $1.0$  & $1.0$  & Same as IC-6 but with no SMBH \\
\hline
\end{tabular} 
\end{center}
\end{table*}
IC-1 - IC-3 represent the progenitor galaxies for the initial 1:1 major merger, adopting the parameters from B16, and differ only by the number of star and DM particles used in the $N$-body realisations. IC-1 is the progenitor at the fiducial resolution, IC-2 has the same stellar resolution as IC-1 but has ten times more dark matter particles,
and IC-3 has the same dark matter resolution as IC-1 but with ten times more star particles. The stellar mass is set by assuming that the observed mass is already present in the progenitor galaxies of the initial major merger. The effective radius is set by assuming that it increases by a factor of two in a dry equal mass merger \citep{Naab2009}, as a consequence of the virial theorem. IC-4 - IC-6 correspond to the progenitor galaxies for the subsequent minor and major mergers: in IC-4 the structural parameters are obtained via abundance matching \citep{behroozi13} assuming an effective radius $R_{\rm {e}}=0.015\,r_{200}$ \citep{kravtsov2013}, IC-5 is a scaled down model of IC-1 by a factor of 5 and IC-6 a scaled down model of IC-1 by a factor of 2. IC-6 - IC-8 represent the progenitor galaxies for the initial 1:1 major merger assuming the parameters by D19: IC-7 adopts the SMBH mass from B16 while IC-8 adopts the SMBH mass from D19.  The merger parameters are given in Table \ref{tab:merger_time_params}.
\begin{table*}
\begin{center}
\caption{Merger parameters. Runs A1-3 correspond to the first equal-mass merger. Runs B1-4 are subsequent mergers from remnant A1. Run C1 is a subsequent 1:3 merger from B4. Run D1 is a merger simulation using the S\'ersic parameters from D19 but with SMBH mass estimated from B16. Run D2 is a merger simulation using S\'ersic parameters and SMBH mass derived from the  $R_{b}-M_{\bullet}$ relation by D19. Columns from left to right: merger remnant label, primary ($M_{\bullet1}$) and secondary ($M_{\bullet2}$) galaxy SMBH mass, total dark-matter ($\mdm$) and stellar ($M_{\star}$) mass, number of stellar ($N_{*}$) and dark-matter ($\ndm$) particles, mass ratio of the galaxy merger ($q$), initial separation ($r_{\rm {sep}}$) and orbital eccentricity ($e$) of the progenitor centres, labels of the progenitor models. }
\label{tab:merger_time_params}
\begin{tabular}{c c c c c c c c c c c} 
\hline 
\hline
Remnant & $M_{\bullet1}$ & $M_{\bullet2}$ & $\mdm$ & $M_{*}$ & $N_{*}$ & $\ndm$ & $q$ & $r_{\rm {sep}}$ & $e$ & Progenitors \\[0.8ex]
      & [$10^{10}\msun$]  & [$10^{10}\msun$]  & [$10^{10}\msun$] & [$10^{10}\msun$] & [$10^{6}$] & [$10^{6}$] & & [$\kpc$] & &  \\
\hline 
A1  &  $1.27$ &  $1.27$ & $2.0\times10^{4}$ &  $4.4\times10^{2}$ & $2.0$ & $2.0$ & $1.0$ & $39$ & $0.95$ & IC-1 + IC-1 \\
A2  &  $1.27$ & $1.27$ & $2.0\times10^{4}$ &  $4.4\times10^{2}$ & $2.0$ & $20.0$ & $1.0$ & $39$ & $0.95$ & IC-2 + IC-2
\\
A3  &  $1.27$ & $1.27$ & $2.0\times10^{4}$ &  $4.4\times10^{2}$ & $20.0$ & $2.0$ & $1.0$ & $39$ & $0.95$ & IC-3 + IC-3 \\
A1E  &  $0$ &  $0$ & $2.0\times10^{4}$ &  $4.4\times10^{2}$ & $2.0$ & $90.0$ & $1.0$ & $39$ & $0.95$ & IC-1aE + IC-1aE \\
\hline
B1  &  $2.54$ & $0.101$ & $2.2\times10^{4}$ &  $4.6\times10^{2}$ & $3.0$ & $3.0$ & $0.1$ & $45$ & $0.95$ & A1 + IC-4  \\
B2  &  $2.54$ & $0.254$ & $2.2\times10^{4}$ &  $4.8\times10^{2}$ & $3.0$ & $3.0$ & $0.1$
& $45$ & $0.95$ & A1 + IC-5 \\
B3  &  $2.54$ & $0.635$ & $2.5\times10^{4}$ &  $5.5\times10^{2}$ & $3.0$ & $3.0$ & $0.25$ & $45$ & $0.95$ & A1 + IC-6 \\
B4  &  $2.54$ & $1.27$ & $3.0\times10^{4}$ &  $6.6\times10^{2}$ & $3.0$ & $3.0$ & $0.5$ & $45$ & $0.95$ & A1 + IC-1 \\
B3a  &  $2.54$ & $0$ & $2.5\times10^{4}$ &  $5.5\times10^{2}$ & $3.0$ & $3.0$ & $0.25$ & $45$ & $0.95$ & A1 + IC-6a \\
B4a  &  $2.54$ & $0$ & $3.0\times10^{4}$ &  $6.6\times10^{2}$ & $3.0$ & $3.0$ & $0.5$ & $45$ & $0.95$ & A1 + IC-1a \\
\hline
C1  &  $3.81$ & $1.27$ & $4.0\times10^{4}$ &  $8.8\times10^{2}$ & $4.0$ & $4.0$ & $0.33$ & $45$ & $0.95$ & B4 + IC-1 \\
\hline
\hline
D1  &  $1.27$ &  $1.27$ & $2.0\times10^{4}$ &  $4.4\times10^{2}$ & $2.0$ & $2.0$ & $1.0$ & $20$ & $0.95$ & IC-7 + IC-7  \\
D2  &  $3.23$ &  $3.23$ & $2.0\times10^{4}$ &  $4.4\times10^{2}$ & $2.0$ & $2.0$ & $1.0$ & $20$ & $0.95$ & IC-8 + IC-8 \\
\hline
\end{tabular} 
\end{center}
\end{table*}
The merger remnants A1-A3 are from the initial equal mass major merger with different resolutions. Merger remnants B1-B4 correspond to subsequent dry mergers after the initial merger of A1. B1 represents a 1:10 minor merger between the merger remnant A1 and IC-4, where the secondary galaxy has structural parameters determined from abundance matching \citep{behroozi13}. B2 represents a 1:10 minor merger between the merger remnant A1 and IC-5, where the secondary galaxy is a scaled down progenitor from A1. B3 represents a 1:4 minor merger between the merger remnant A1 and IC-6, where the secondary galaxy is scaled down from the remnant A1. B4 represents a 1:2 major merger between the merger remnant A1 and IC-1, where the secondary galaxy is the progenitor of the initial 1:1 major merger. C1 is the subsequent 1:3 major merger between remnant B4 and IC-1, where the secondary galaxy is the progenitor of the initial 1:1 major merger. D1 corresponds to the 1:1 initial major merger at the fiducial resolution using the S\'ersic parameters from D19 but with the SMBH mass from B16. D2 corresponds to the 1:1 initial major merger at the fiducial resolution using the S\'ersic parameters from D19 and the SMBH mass derived from the $R_{\rm {b}}-\msbh$ from D19. Prior to all the 
subsequent mergers (B1-B4 and C1) we merge the binary black holes into a single SMBH located at the binary's centre of mass: A1 for mergers B1-B4 and B4 for the merger C1.\newline
For the initial 1:1 merger (A1-A3 and D1-D2) the galaxies were placed at a distance greater than $2R_{\rm {e}}$ of the progenitor\footnote{We verified that this distance was sufficient by repeating mergers A1-A3 with separations of $100\kpc$ and $250\kpc$, and we observed no significant difference in the evolution.} and on a bound elliptical orbit with eccentricity $e=0.95$\footnote{We define the eccentricity of the orbit by assuming that the galaxies are point masses. We note that this is an approximation due to the overlap between the progenitor galaxies.}.
Such a large eccentricity was chosen to mimic merger conditions in cosmological simulations \citep[e.g.][]{khochfar2006eccen} as well as to reduce computational time.

\subsection{Numerical method}
\label{sec:numerical}
We perform all merger simulations with the code \textsc{griffin} \citep{dehnen2014griffin}, which uses the Fast Multiple Method as force solver for gravity between stars and dark matter, using adaptive parameters meant to avoid a tail of large force errors, with a mean relative force error of $3\times10^{-4}$. SMBH gravity is computed by direct summation and all trajectories are integrated using the leapfrog integrator. \textsc{griffin} has recently been shown to evolve BHBs in galactic mergers as accurately as direct summation methods \citep{nasim2020,Nasim2021}. For all of our numerical simulations, we adopt a softening length $\epsilon = 23\pc$ for the stars,  dark matter particles and the SMBHs. We verified that this choice of softening is appropriate by repeating merger A1 with a softening value $\epsilon= 2.3\pc$. The dynamical friction phase is unaffected while the three-body phase shows a slightly reduced central surface brightness, though the difference is insignificant. We use a softening kernel that corresponds to smooth source particles with density proportional to $(r^2+\epsilon^2)^{-7/2}$ rather than the more commonly used Plummer softening which is characterised by a density proportional to $(r^2+\epsilon^2)^{-5/2}$. For a fixed value of $\epsilon$ the source mass is more concentrated than for the Plummer case, implying a weaker softening.
The SMBH to star particle mass ratio for the initial conditions is in the range $5720 \leq \msbh / m_{\star} \leq 57200$ while the SMBH to DM particle ratio is in the range $127 \leq \msbh / m_{\rm{DM}} \leq 1270$ (see Table~\ref{tab:merger_time_params}), 
ensuring reduced stochastic effects in the evolution of the BHB \citep{nasim2020}.

\section{Initial merger}
\label{sec:firstmerger}
We start our analysis by investigating the effects of the first merger, i.e. the major mergers A1-A3 and D1-D2, focusing on the evolution of the BHB that forms and the formation of a core in the stellar density profile.

\subsection{Binary scouring}
\label{sec:scouring}
\begin{figure*}
    \centering
    \includegraphics[width=2.0\columnwidth]{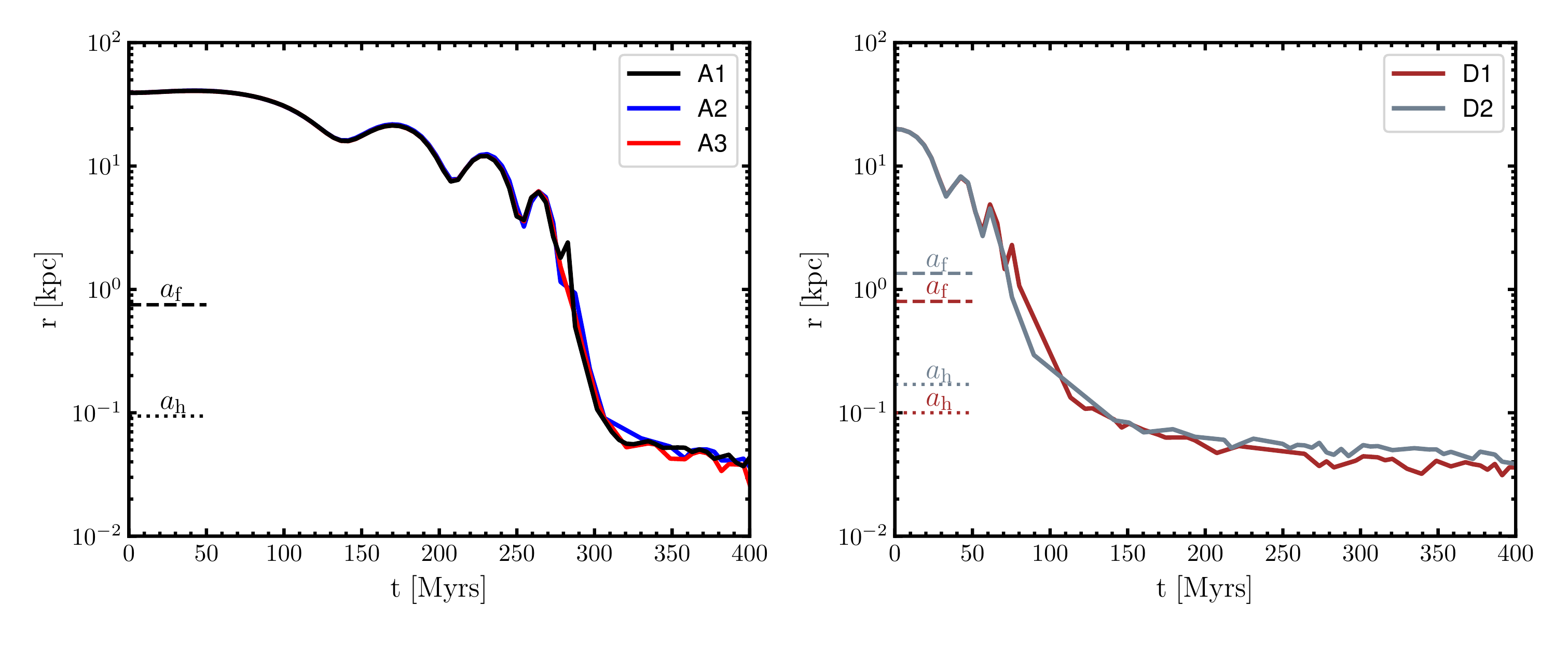}
    \caption{Distance between the SMBHs as a function of time for the 1:1 initial major mergers: A1-A3 (left panel) and D1-D2 (right panel).
    The horizontal dashed/dotted lines correspond to the $\af$/$\ah$ separations.}
    \label{fig:major_sep}
\end{figure*}

\begin{figure*}
    \centering
    \includegraphics[width=2.0\columnwidth]{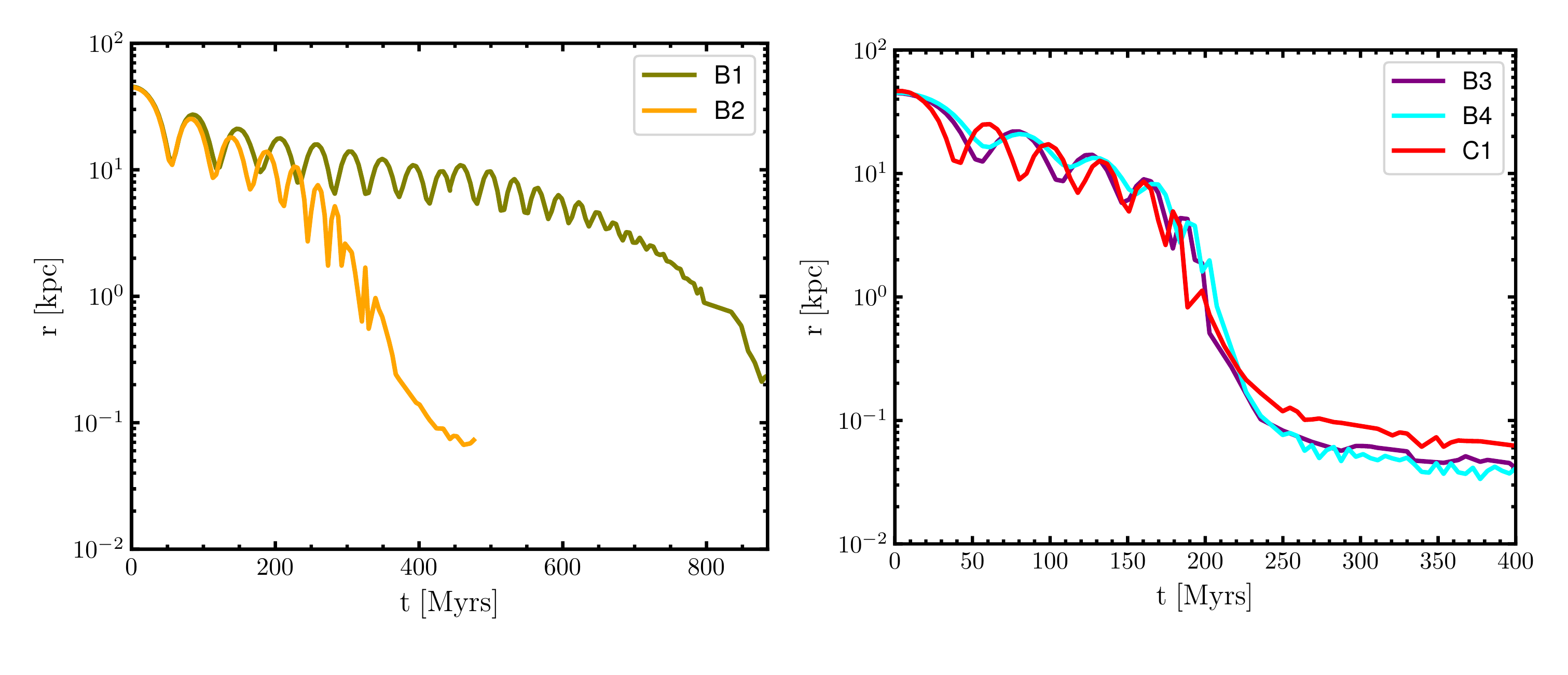}
    \caption{Distance between the SMBHs as a function of time for the subsequent mergers following the initial 1:1 major merger: B1-B2 (left panel) and B3-C1 (right panel). }
    \label{fig:subsequent_sep}
\end{figure*}

When two progenitor galaxies merge, the SMBHs sink to the centre of the merger remnant via the process of dynamical friction \citep{chandrasekhar1943} from the surrounding stellar and dark matter distribution. This process drives the SMBHs to a separation $\af$ at which the stellar mass $M_{\star}$ within the binary orbit is equal to twice the mass of the secondary black hole
\begin{equation}
    M_{\star}\left( r<\af \right) = 2 M_2.
\end{equation}
The time when the binary reaches $\af$ roughly corresponds to the time when the SMBHs become gravitationally bound  and is commonly taken to mark the transition from the dynamical friction dominated phase to the hardening dominated phase.  This occurs at $t\approx 287\myr$ in remnants A1-A3 and $t\approx 80\myr$ in remnants D1 and D2, as can be seen in Fig.~\ref{fig:major_sep} and \ref{fig:subsequent_sep} where we show the distance between the SMBHs as a function of time for the initial 1:1 major mergers (Fig.~\ref{fig:major_sep}; simulations A1-A3 and D1-D2) and all subsequent mergers (Fig.~\ref{fig:subsequent_sep}; simulations B1-B2 and B3-C1).\newline
Rapid hardening ensues as three-body encounters with stars become important and extract energy and angular momentum from the orbit, causing both a shrinking in the binary separation and ejection of stars to larger distances. As the central density is lowered and a core is a carved due to stellar ejections, this process is generally called {\it binary scouring}. The binary is formally considered hard at the 
{\it hard binary} separation $\ah$, where the specific binding energy of the BHB exceeds the average specific kinetic energy of the stars \citep{merritbook2013}
\begin{equation}
    \ah = \frac{G\mu}{4\sigma^2}
\label{eq: hard_binary_sep_sigma}
\end{equation}
with $\mu = M_1M_2/(M_1+M_2)$ the reduced mass and $\sigma$ the line of sight velocity dispersion. An alternative definition which is better suited to $N$-body simulations is given by 
\begin{equation}
    \ah = \frac{\mu}{M_{\rm bin}}\frac{\rmo}{4} = \frac{q}{(1+q)^2} \frac{\rmo}{4}
\label{eq: hard_binary_sep}
\end{equation}
where $M_{\rm bin}$ is the mass of the BHB, $q = M_2/M_1$ is the black hole mass ratio and $\rmo$  represents the radius containing a mass in stars equal to twice the mass of the primary black hole. We adopt this second definition, which for equal mass binaries simply gives $\ah = \rmo/16$.

\begin{figure*}
    \centering
    \includegraphics[width=2.2\columnwidth]{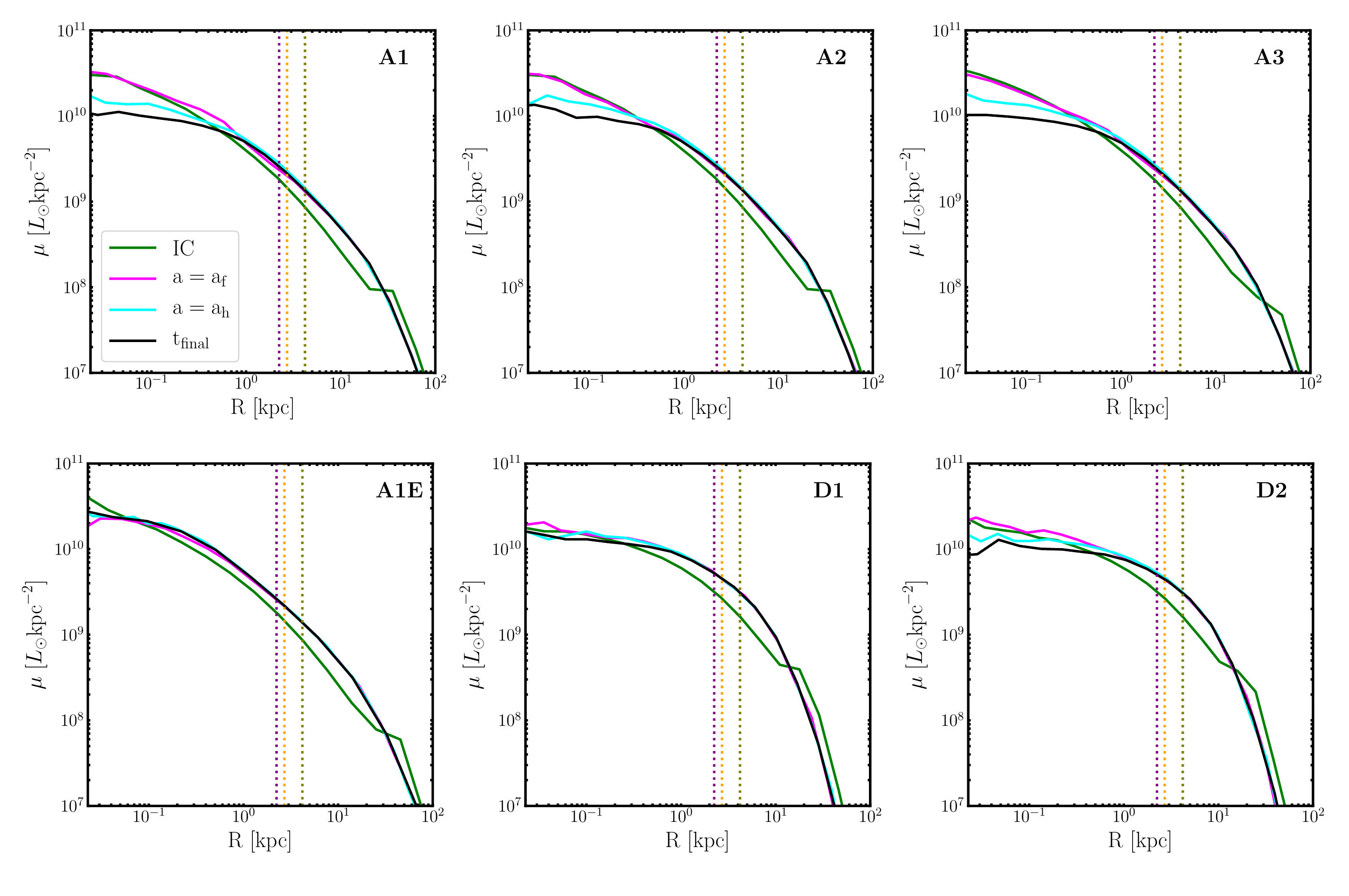}
    \caption{Surface luminosity profiles for the initial equal mass major mergers (A1-3, top panels),  the equal mass major merger without SMBHs and with equal mass star and DM particles (A1E, bottom left panel) and the equal mass major mergers using D19 parameters: D1 and D2  (bottom left and bottom right panel respectively). The profiles are computed at different times during the evolution: the initial condition (green line), the times when $\af$ (magenta) and $\ah$ (cyan) are reached and the end of the $N$-body integration (black). The vertical dotted lines represent the three largest core radii known to date: IC 1101 (olive) with a $4.2\kpc$ core, A2261-BCG (orange) with a $2.71\kpc$ core and 4C +74.13 (purple) with a $2.24\kpc$ \citep{dullo2019}. The innermost value of the radial range is set by the softening length of the stellar particles. Initial equal mass mergers with SMBHs show evidence of binary scouring as well as as violent relaxation at larger radii, while no core is formed in the simulation without SMBHs. The surface mass densities can be computed by multiplying the surface luminosity by $\Upsilon=3.5M_\odot/L_\odot$.}
    \label{fig:initial_merger_surface_profiles}
\end{figure*}

\begin{figure*}
    \centering
    \includegraphics[width=2.2\columnwidth]{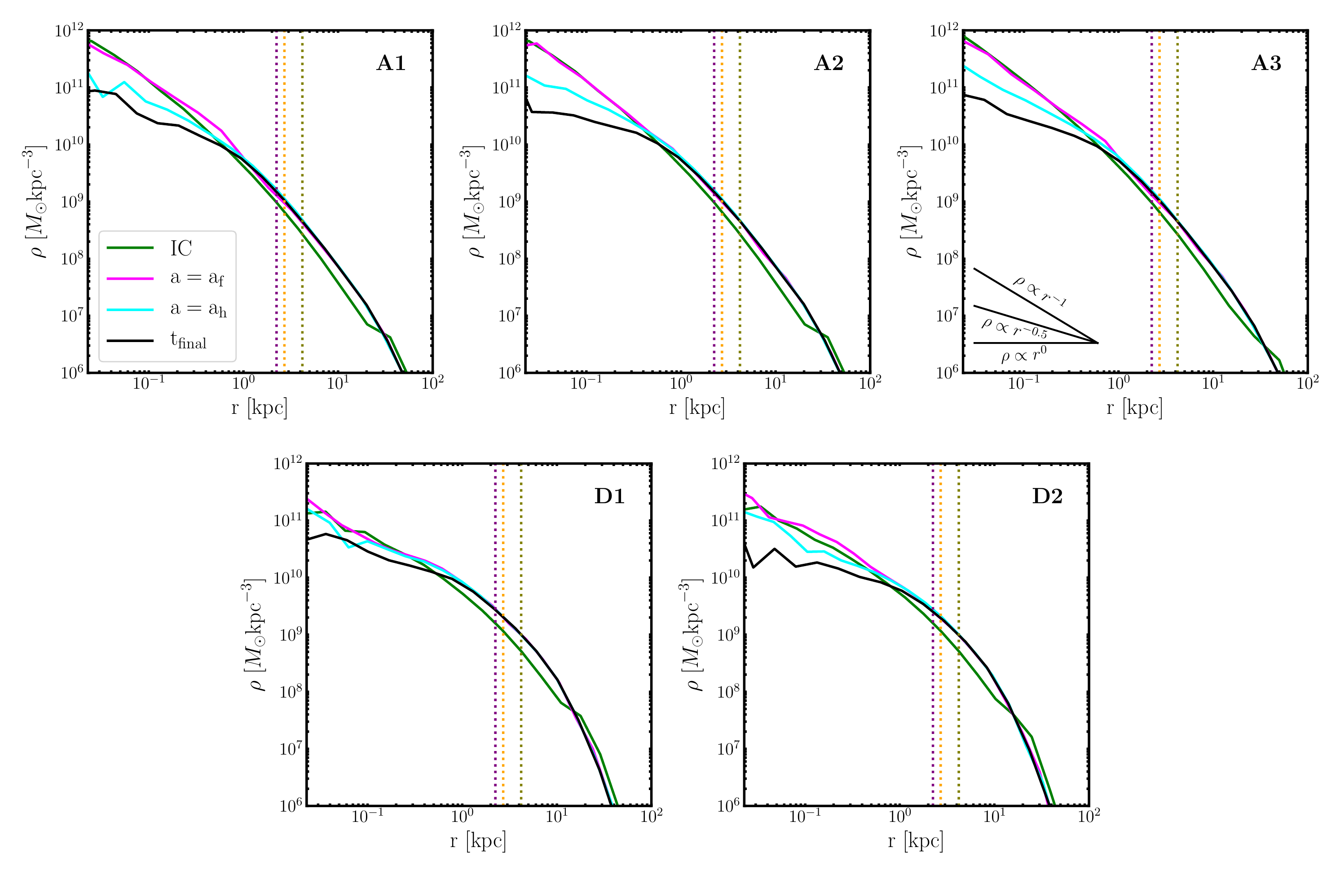}
    \caption{Spatial density profiles for the initial equal mass major mergers with SMBHs (A1-3, top panels) and the equal mass major mergers using D19 parameters: D1 and D2  (bottom left and bottom right panel respectively). The vertical dotted lines are as defined in Fig.~\ref{fig:initial_merger_surface_profiles}.
    Initial equal mass mergers with SMBHs form a weak central cusp in spatial density. }
    \label{fig:initial_merger_density_profiles}
\end{figure*}

We find $\af=0.75\kpc$ and $\ah = 0.094\kpc$ for runs A1-A3 and
$\af=0.80,1.35\kpc$, $\ah=0.10,0.17\kpc$ for runs D1 and D2, respectively. The larger values for the D1-D2 remnants may be surprising, considering that the progenitor galaxies have a smaller effective radius $R_{\rm {e}}$ (see Table \ref{tab:core_sersic_params}). However this can be attributed to the fact that the progenitor profiles in both D1-D2 have a significantly smaller S\'ersic index than those in A1-A3, 
meaning they are less cuspy. The resulting lower central density
can also be seen in the initial condition profiles in Fig.~\ref{fig:initial_merger_surface_profiles}. 
In addition, the SMBH mass is larger in the D2 progenitor model, implying a larger radius of influence and hard binary separation.
We run all of our merger simulations past the hard binary separation but end the simulation before the force softening length is reached.
We show the surface luminosity profiles of models A1-A3 and D1-D2 in Fig.~\ref{fig:initial_merger_surface_profiles} (upper panels and lower middle/right panels respectively). 
All profiles are computed with respect to the primary SMBH, which represents a viable centre even at early times \footnote{ We have verified that adopting different projections, as appropriate for merger remnants, yields no significant differences in the surface luminosity profiles.}.
As a reference, the vertical dotted lines in the figures indicate the three largest core radii known to date: IC 1101 (olive) with a $4.2\kpc$ core, A2261-BCG (orange) with a $2.71\kpc$ core and 4C+74.13 (purple) with a $2.24\kpc$ \citep{dullo2019}.\newline
The profiles of remnants A1-A3 show clear evidence of binary scouring: the central surface luminosity decreases during the binary hardening phase, between the end of the merger (marked roughly by the time when the separation $\af$ is reached) and the time when the binary reaches the hard-binary separation. This is due to the ejection of stars from the central region as a result of three-body interactions \citep[e.g.][]{quinlan1996,milosavljevicmeritt2001,merritt2006}. The ejected mass is deposited at larger distances but it is not clearly visible in the profiles as it is spread out over a large volume. No significant difference is found due to the adopted resolution in stellar and DM particles, as can be seen comparing models A2 and A3 with the fiducial resolution model A1.\newline
We do not observe evidence of significant scouring in model D1 (See Fig.\ref{fig:initial_merger_surface_profiles}, bottom middle panel), while some scouring is present in model D2 (See Fig.\ref{fig:initial_merger_surface_profiles}, bottom right panel). This can be attributed to the flatter central profiles of the D1-D2 progenitors as compared to the A1-A3 progenitors, as a lower central density leads to reduced binary scouring \citep[e.g.][]{rantala2018}. In D2, this effect is, however, offset by its more massive SMBH, leading to significant scouring  (see Fig.~\ref{fig:initial_merger_surface_profiles}, bottom right panel).\newline
To verify that binary scouring is the only mechanism responsible for the central core, we also ran an equal mass merger simulation without SMBHs (A1E). We adapted the resolution to have equal mass star and DM particles which ensures there are no coring effects due to mass segregation \footnote{Mass segregation effects would be present at the fiducial resolution, leading to a small inner core due to the outward migration of lower mass particles. Such effects are not seen in the simulations with SMBHs (A1-A3) due to the effects of binary scouring.}. The surface luminosity profile is given in Fig.~\ref{fig:initial_merger_surface_profiles} (bottom left panel), centred on the primary galaxy \citep{power2003} and evaluated at the same times considered for the simulation with SMBHs (A1). As expected, no core is formed in the model without SMBHs.\newline
We note that in this work we adopt the observational approach of defining a core as a flattening in the surface luminosity profile, measured by the break radius of the best fitting core-S{\'e}rsic model (see section \ref{sec:core}). However, a core in the surface density profile doesn't necessarily imply a core in the spatial density profile; a shallow cuspy profile would also appear cored in projection. The spatial density profiles for the initial major mergers (shown in Fig.~\ref{fig:initial_merger_density_profiles})
reveal indeed the presence of a shallow cusp in all cases, scaling approximately as $\rho(r) \sim r^{-\gamma}$, with $0.5<\gamma < 1$. As we will show in Section \ref{sec:GW_section}, cores produced by GW recoil of a SMBH are, however, of a different nature, and correspond to non-divergent profiles in spatial density.

\subsection{Violent relaxation}
\label{sec:mixing}
The time evolving gravitational potential of a galactic major merger results in violent relaxation and dynamic mixing in which stars experience energy changes that are dependent upon their initial positions and velocities. This mixing operates on the timescale of the merger itself, and is therefore important at early times. Evidence can be seen in the surface luminosity profiles of mergers D1 and D2 in the form of deposition of material at intermediate radii. A a more modest effect is seen in merger A1, due to its higher stellar density.\newline 
\begin{figure*}
    \centering
    \includegraphics[width=2.0\columnwidth]{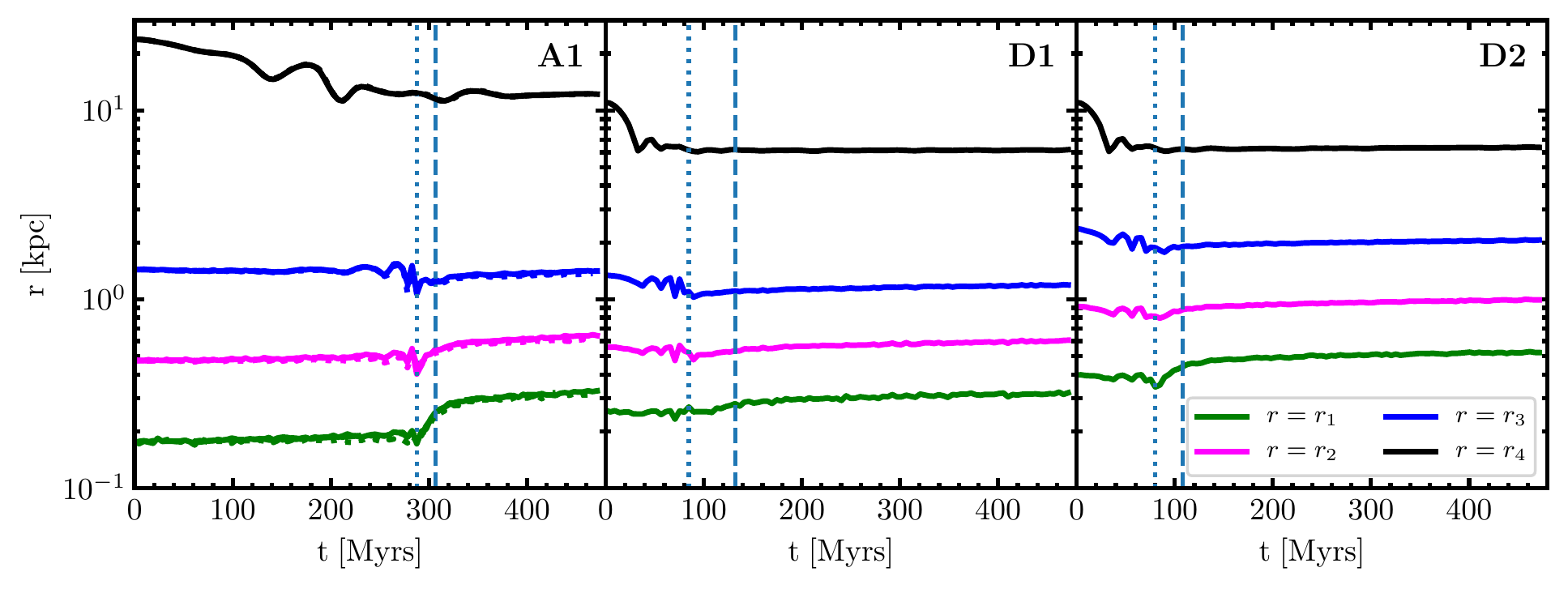}
    \includegraphics[width=2.0\columnwidth]{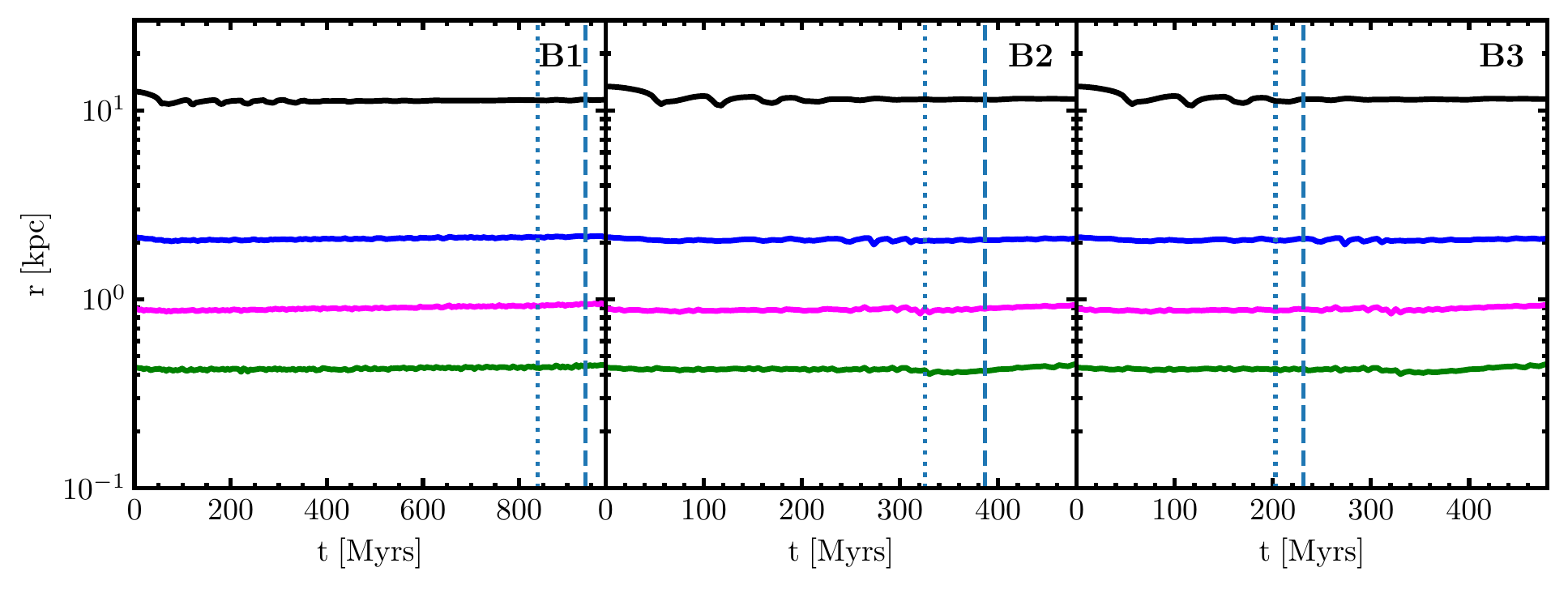}
    \includegraphics[width=1.4\columnwidth]{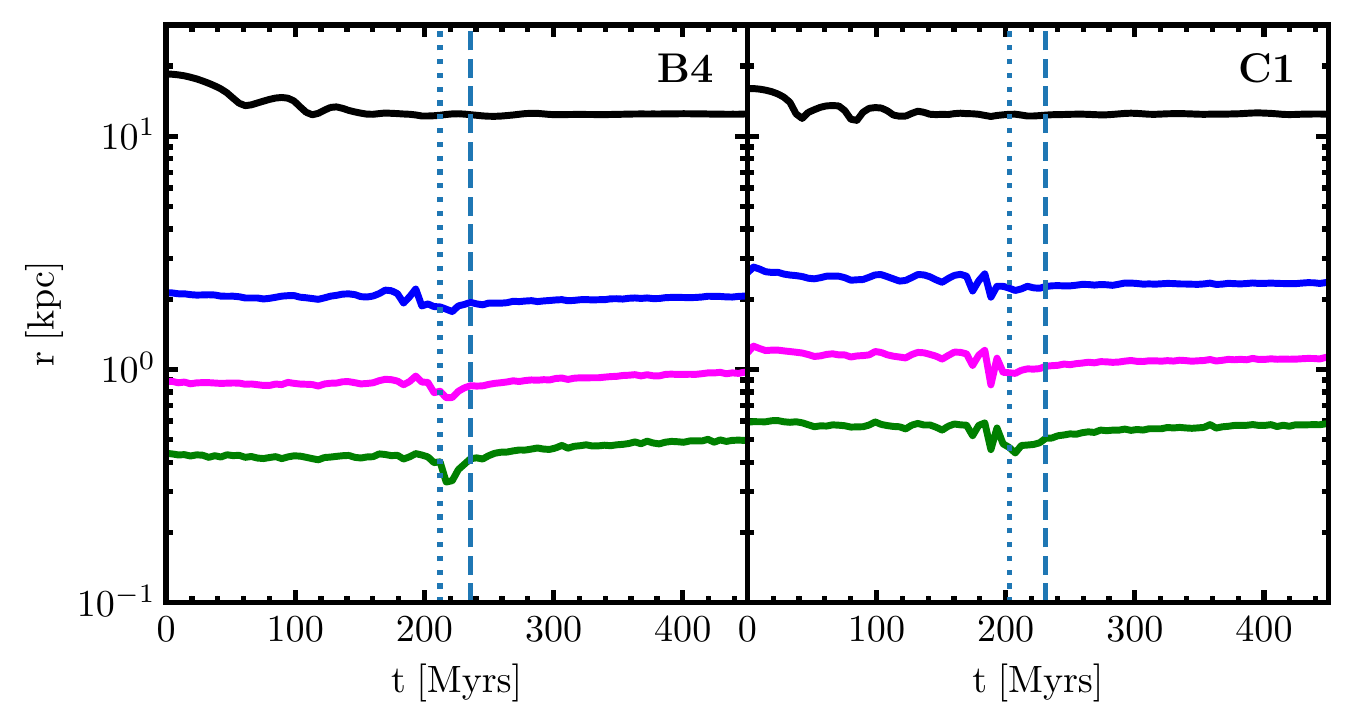}
    \caption{Evolution of the Lagrangian radii defined in Eq.~\ref{eq:r_S_radii} for the different merger remnants. The Lagrangian radii for merger runs A2 and A3 are represented by dotted/dashed lines in the upper left panel which approximately lie on top of the Lagrangian radii for model A1. The vertical dotted/dashed lines correspond to the times when $a=\af$ and  $a=\ah$ are reached, respectively. Both binary scouring and tidal deposition can be clearly seen in the Lagrangian radii. Scouring  is particularly effective in the equal mass merger A1, whereas tidal deposition is significant in models D1 and D2, as well as  in subsequent mergers. The time evolution also shows that tidal deposition operates at early times, while the merger is ongoing, while scouring requires a SMBH pair and operates after the end of the dynamical friction phase.}
    \label{fig:lagrange_radii_A1_D1_D2}
\end{figure*}
The effects of binary scouring and violent relaxation can be seen also in the evolution of the Lagrangian radii. We consider four radii corresponding to the following values of enclosed stellar mass, centred on the primary SMBH:
\begin{subequations}
	\begin{align}
	\label{eq:r_s1}
	&M_{\star}\left(r<r_1\right) = \tfrac15 \msbh\\
	\label{eq:r_s2}
	&M_{\star}\left(r<r_2\right) = \msbh\\
	\label{eq:r_s3}
	&M_{\star}\left(r<r_3\right) = 5\msbh\\
	\label{eq:r_s4}
	&M_{\star}\left(r<r_4\right) = \tfrac14 M_{\star}.
	\end{align}
	\label{eq:r_S_radii}
\end{subequations}
The combined stellar mass of both progenitor galaxies is considered, so at times when the galaxies are well separated the radius $r_4$ approximates the half mass radius of the progenitor during a 1:1 equal mass merger. The evolution of the Lagrangian radii is shown in Fig.~\ref{fig:lagrange_radii_A1_D1_D2} for the different models. In model A1 the Lagrangian radius $r_4$ decreases throughout the evolution, with the dips corresponding to the pericentre passages of the galaxies.  A qualitatively similar behaviour is seen in models D1 and D2, but with a faster initial drop and less pronounced dips at pericentre passages due to the much faster inspiral of the secondary galaxy in these models. The innermost radius $r_1$ is approximately constant in model A1 during the merger and increases during binary hardening, as a result of the stellar ejections from the binary. Binary scouring is most effective prior to the binary reaching $\ah$, and then continues at a lower rate, as also seen in the simulations of \citet{rantala2018}. On the other hand, $r_1$ increases very slowly in the D1 remnant, supporting the earlier conclusion that binary scouring is inefficient in this model. The evolution of $r_1$ in model D2 is intermediate between that of models A1 and D1, again supporting the conclusion of a modest binary scouring mediated by the more massive central SMBH.

\section{Subsequent mergers}
\label{sec:multiple_mergers}

\begin{figure*}
    \centering
    \includegraphics[width=2.2\columnwidth]{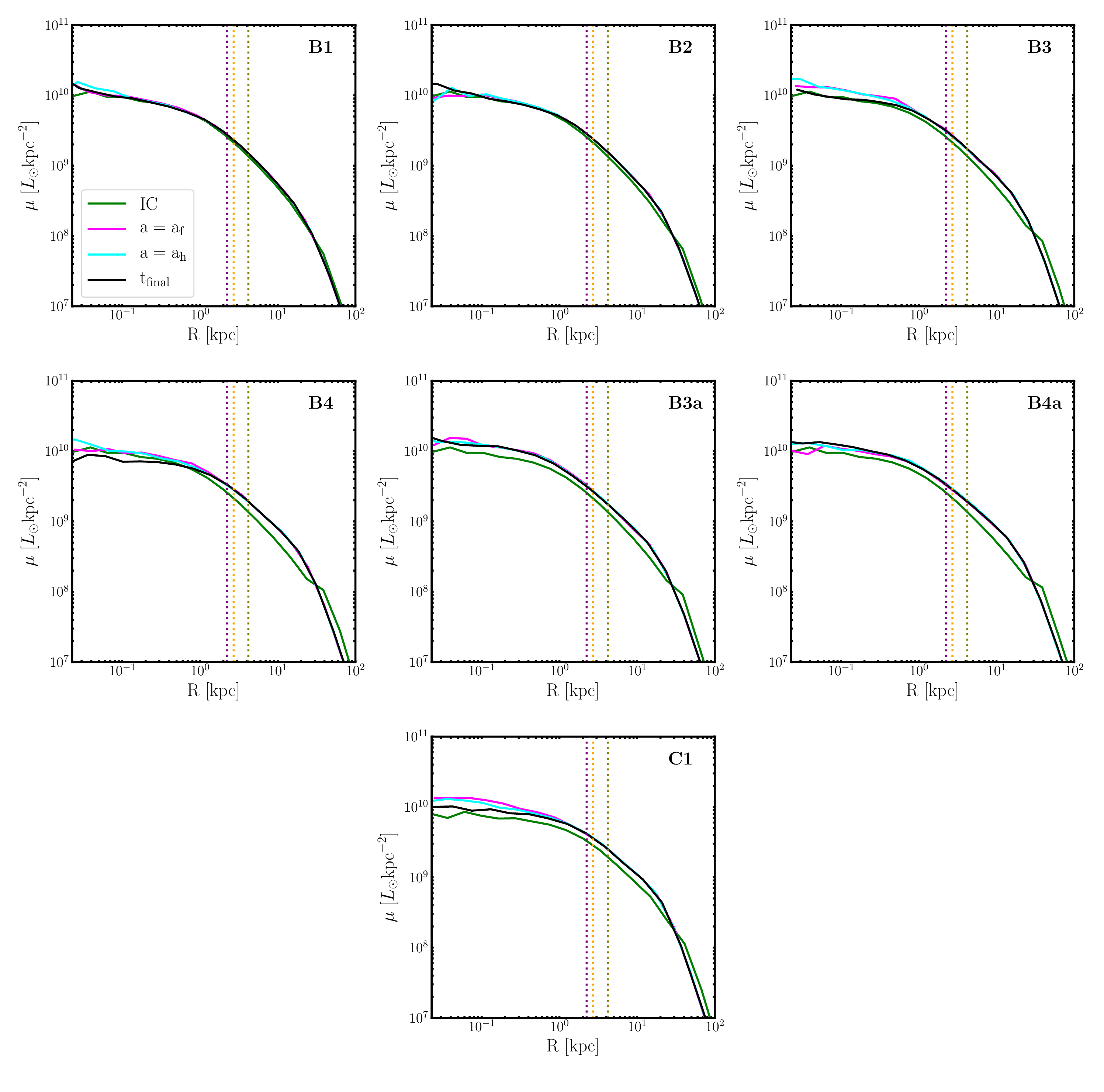}
    \caption{Surface luminosity profiles for the subsequent mergers: (B1-3, top panels), 1:2 major merger (B4, middle left panel), 1:4 and 1:2 mass mergers without secondary SMBH: B3a and B4a  (central and middle right panel respectively), 1:3 mass merger subsequent from B4 (C1, bottom panel).
    The profiles are computed at different times during the evolution: the initial condition (green line), the times when $\af$ (magenta) and $\ah$ (cyan) are reached and the end of the $N$-body integration (black). The vertical dotted lines represent the three largest core radii known to date: IC 1101 (olive) with a $4.2\kpc$ core, A2261-BCG (orange) with a $2.71\kpc$ core and 4C +74.13 (purple) with a $2.24\kpc$ \citep{dullo2019}. 
    Subsequent mergers with SMBHs primarily increase the size of the core via further tidal deposition. The mass deficit carved in the initial equal mass major merger does not grow linearly with the number of subsequent mergers. The subsequent mergers without BHBs increase their central concentration to absence of binary scouring. The surface mass densities can be computed by multiplying the surface luminosity by $\Upsilon=3.5M_\odot/L_\odot$.}
    \label{fig:subsequent_merger_surface_profiles}
\end{figure*}

\begin{figure*}
    \centering
    \includegraphics[width=2.2\columnwidth]{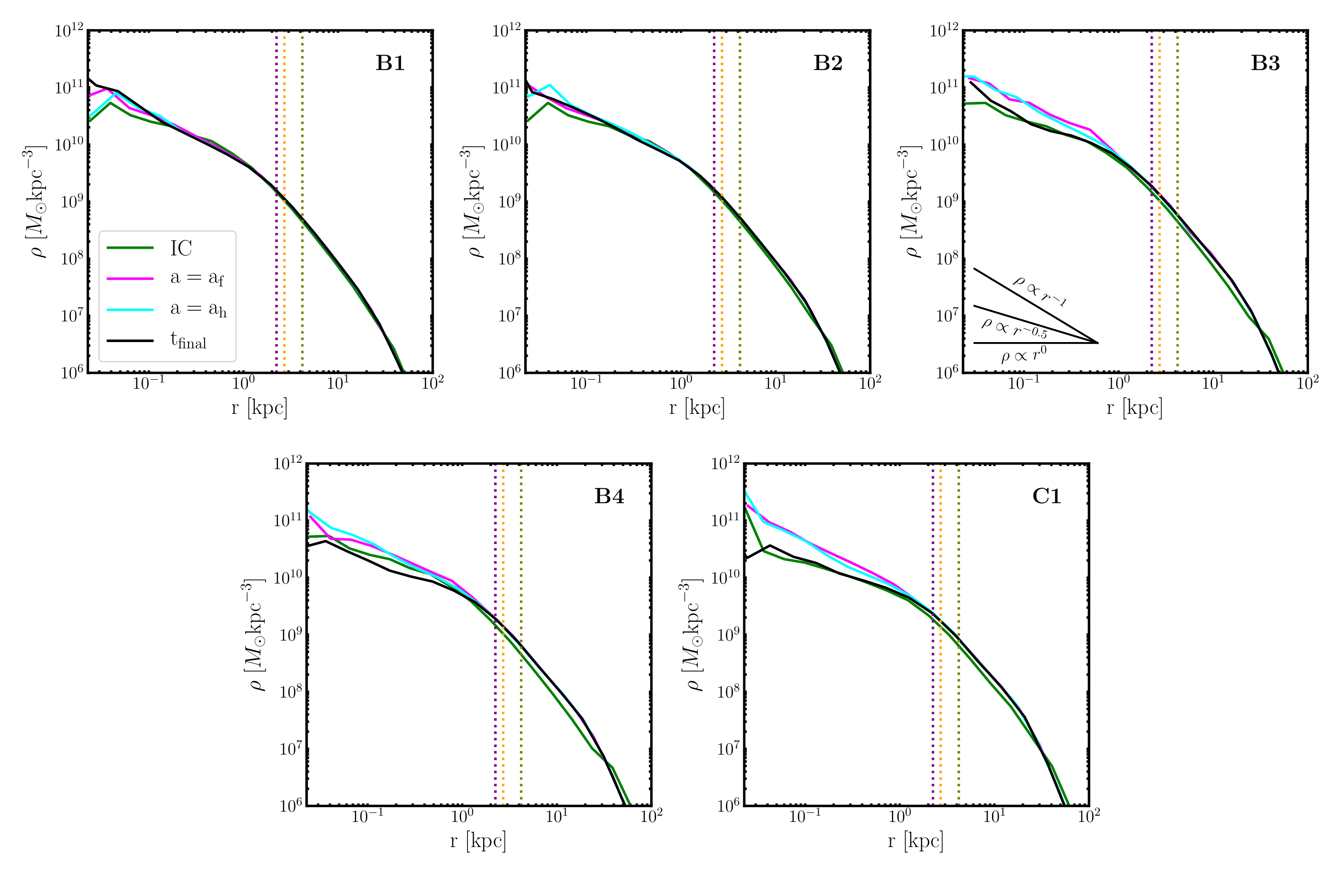}
    \caption{Spatial density profiles for the subsequent mergers: (B1-3, top panels), 1:2 major merger (B4, lower left panel) and 1:3 mass merger subsequent from B4 (C1, bottom right panel).  The vertical dotted lines are as defined in Fig.~\ref{fig:initial_merger_surface_profiles}. The spatial density profiles for the subsequent mergers show evidence for the formation of a weak central cusp which appears cored in projection.}
    \label{fig:subsequent_merger_density_profiles}
\end{figure*}

We now investigate the effects of subsequent mergers following the first 1:1 mass merger A1, which are expected to drive the assembly of ellipticals \citep[e.g.][]{Naab2009,NaabOstriker2017}. We extend the $N$-body integration of model A1 for a further $\sim 200$ $\rm{Myrs}$ to ensure that the surface luminosity profile is no longer evolving and core formation has halted. By this time, the separation between the SMBHs has become comparable to the softening length. We then estimate the coalescence time of the BHB due to the combined effects of three-body scatterings and emission of GWs. We adopt the semi-analytic model, referred to as the Continuous Coefficients Method (CCM), for the evolution of the binary's orbital elements presented in \citet{nasim2020} which includes a time-dependent hardening rate. The CCM approximates the evolution of the BHB  by using a time dependent polynomial extrapolation to model three-body effects and the Post Newtonian prescription by \citet{peters1964} to model the effects of GW emission. The time dependent polynomial extrapolation naturally accounts for different density profiles by fitting the hardening rate of the binary.\newline
We find that the coalescence time for the BHB in model A1 is  $\sim 3.5$ Gyr, short enough that we can safely assume that a sequence of mergers can follow within a Hubble time. We then combined the two SMBHs into one, placed it at the binary's centre of mass and modelled subsequent mergers onto the A1 remnant.\newline
We first consider the B1-B4 mergers where the primary galaxy is the A1 remnant and the secondary is as described in Table \ref{tab:merger_time_params}. While B1 and B2 are 1:10 mergers 
,B3 is a 1:4 merger and B4 a 1:2 merger. The secondary progenitor for B1 is setup using abundance matching while the progenitors for remnants B2-B4 are scaled down versions of A1. 
\footnote{We observe that mass segregation is not present in the numerical simulations with SMBHs, but was present in simulations without SMBHs. This motivated the A1E merger which contained equal mass particles without SMBHs.}
We aim to determine whether cores grow over subsequent mergers, as generally expected for major mergers \citep{merritt2006,delucia2007,Naab2009,NaabOstriker2017}, or if they saturate instead, and what effects minor mergers have on the density profiles. 

\subsection{Tidal deposition}
 During a minor merger a sinking satellite has been shown to heat the primary galaxy, causing the density profile to become shallower or even forming a constant density core \citep{read2006}. Because dynamical friction is suppressed in constant density cores, the satellite stalls at the edge of the core and never reaches the centre of the system \citep{2012ApJ...745...83A,2017ApJ...840...31D}. \citet{Goerdt2010} also show that the core is formed at the radius where the enclosed stellar mass matches the mass of the satellite. Therefore, if a satellite is not dense enough to survive the infall to the centre, mass is deposited at the edge of the core, effectively enlarging its size. This represents a further mechanism of core formation, which we name {\it tidal deposition}, that does not operate by lowering the central density
like binary scouring but rather by increasing the outer density. An additional key difference between the two mechanisms is the time at which they operate: while tidal deposition takes place during the merger, i.e. at early times in the evolution, binary scouring becomes efficient after the dynamical friction phase, when the SMBHs become bound.\newline
We find evidence of tidal deposition in models B2-B4. The surface luminosity profiles of the B models are presented in the upper and middle left panels of Fig.~\ref{fig:subsequent_merger_surface_profiles}.
We observe no significant evolution in the central profiles of B1 and B2, however B2 shows evidence of deposition of mass by the time the separation $\af$ is reached. Model B2 hosts a more massive SMBH and stellar bulge than model B1, and this results in the faster infall seen in Fig.~\ref{fig:subsequent_sep}. The earlier tidal stripping experienced by B2 results in the observed modest deposition at large radii. This can also be seen in the Lagrangian radii (Fig.~\ref{fig:lagrange_radii_A1_D1_D2}), where radius $r_4$ shows a small early decline while the innermost radii show no significant variations. However, the secondary galaxy is not sufficiently massive to affect the primary's profile in either model.\newline
The more massive infallers in models B3-B4 result in a more noticeable tidal deposition at early times: the surface luminosity profiles show an increase in the outer regions by the time $\af$ is reached. There is also a small increase in the central density of the primary at early times which is due to the infall of the secondary galaxy. However, this is erased by binary scouring during the hardening phase, which results in a shallower final profile. Binary scouring plays a role in the 1:2 merger B4, as can be seen in the evolution of the $r_1$ Lagrangian radius in Fig.~\ref{fig:lagrange_radii_A1_D1_D2}, but is negligible in model B3 due to the smaller SMBH mass. In order to verify this interpretation, we re-ran models B3 and B4 without a SMBH in the secondary galaxy (model B3a and B4a, see Table.\ref{tab:merger_time_params}). The surface luminosity profiles (presented in the central middle and right panels of Fig. Fig.\ref{fig:subsequent_merger_surface_profiles}) show evidence of tidal deposition as well as a rise in the central density at early times, similarly to the models with a BHB. However, the central increase persists to late times as binary scouring does not operate in these models. The weak cusp observed in the spatial density profile from the initial major merger persists after the subsequent mergers, as shown in Fig.\ref{fig:subsequent_merger_density_profiles}.\newline
The tidal deposition we observe in the minor mergers following the first equal mass merger is in agreement with  models of core formation in ellipticals galaxies via accretion at late times ($z<3$), in which material is stripped from the secondary and is deposited onto the primary, thus increasing the size of the core. During this phase there is a reduction in the central concentration due to heating and dynamical friction from the initial surviving core \citep[e.g.][]{Elzant2001, Naab2009, NaabOstriker2017}. \newline
We note that tidal deposition is not observed in \citet{rantala2019}, who also investigated the effect of subsequent mergers. The reason for this is not fully understood, but likely owes to differences in the initial conditions.\newline
Finally, we evolved the BHB formed in model B4 to coalescence following the same semi-analytic recipe adopted for model A1 and then proceeded to perform a subsequent 1:3 dry major merger (C1) where the secondary progenitor is IC-1, with parameters as described in Table \ref{tab:merger_time_params}. 
The evolution of the surface luminosity profile of C1 is plotted in Fig.\ref{fig:subsequent_merger_surface_profiles} (bottom panel). Perhaps surprisingly for a 1:3 major merger, we observe a qualitatively similar evolution to remnants B3 and B4, with evidence of tidal deposition at large radii and an increase in central density that persists to late times, with no binary scouring. This clearly indicates that scouring is not effective in a previously carved core, and cores do not necessarily grow through subsequent mergers, as generally assumed in theoretical models \citep{merritt2006,rantala2019}. For example, \citet{merritt2006} suggest that core size may be used as an indication of the number of major mergers experienced by a given galaxy. We find no evidence of this behaviour, and this is due to the fact that scouring is not effective in an already depleted core, carved by a previous merger.  
However, if cores saturate in flat profiles, independent mechanisms need to be invoked to explain large cores, with a mass deficit much larger than the assumed SMBH mass.

\subsection{Stalled satellites}
\label{sec:stalled_sat}

\begin{figure*}
    \centering
    \includegraphics[width=2.0\columnwidth]{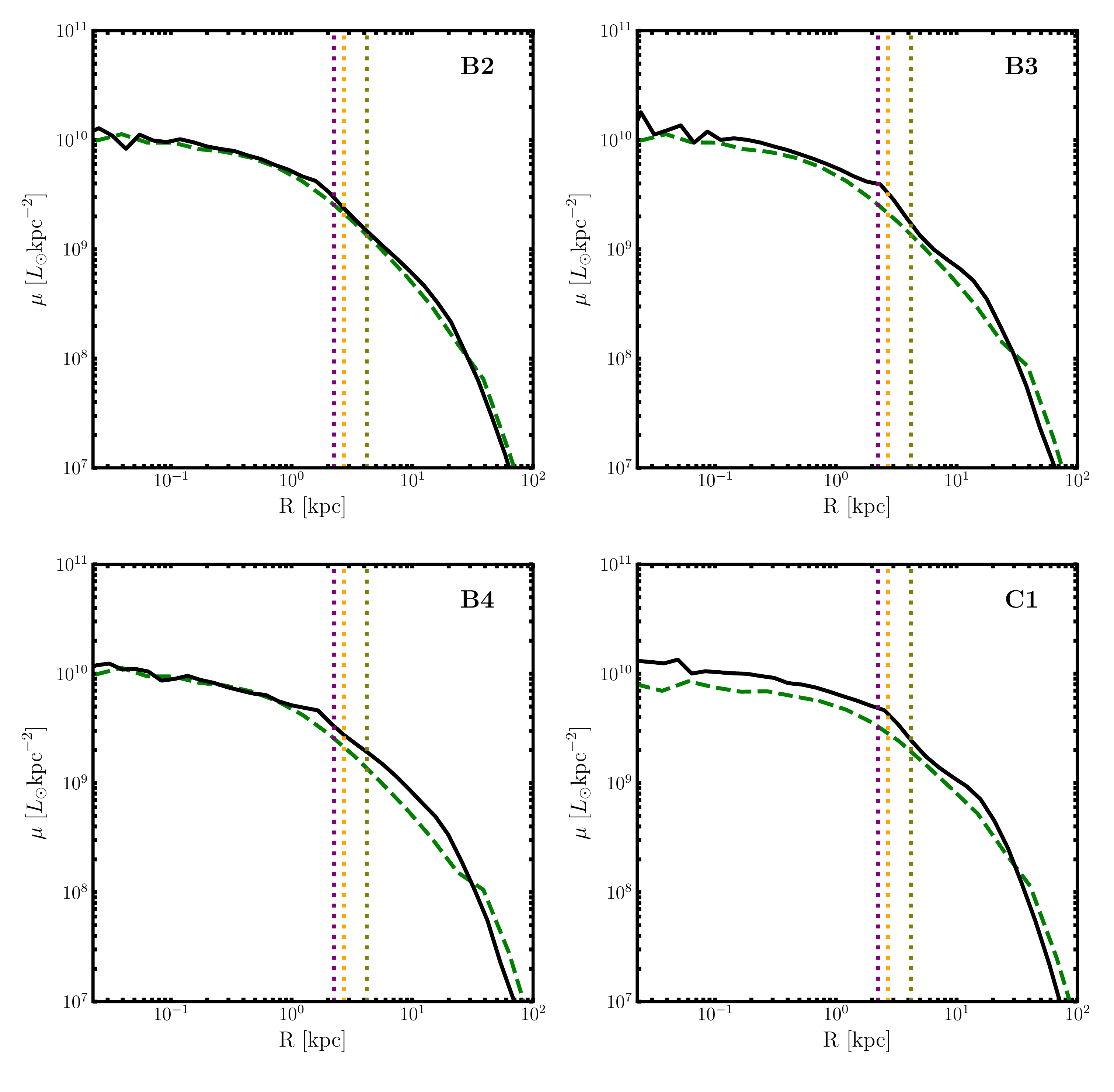}
    \caption{Surface luminosity profiles of subsequent mergers at times showing the formation of bumps (solid lines) to be compared against the initial profiles (dashed lines).
    As in Fig.\ref{fig:initial_merger_surface_profiles}, the vertical dotted lines  represent the three largest core radii known to date: IC 1101 (olive) with a $4.2\kpc$ core, A2261-BCG (orange) with a $2.71\kpc$ core and 4C +74.13 (purple) with a $2.24\kpc$ \citep{dullo2019}. The observed bumps are due to the infall of satellites during minor mergers, and are observed in A2261-BCG \citep{postman2012,bonfini2016}. The surface mass densities can be computed by multiplying the surface luminosity by $\Upsilon=3.5M_\odot/L_\odot$.}
    \label{fig:subsequent_bump}
\end{figure*}

\begin{figure*}
    \centering
    \includegraphics[width=2.0\columnwidth]{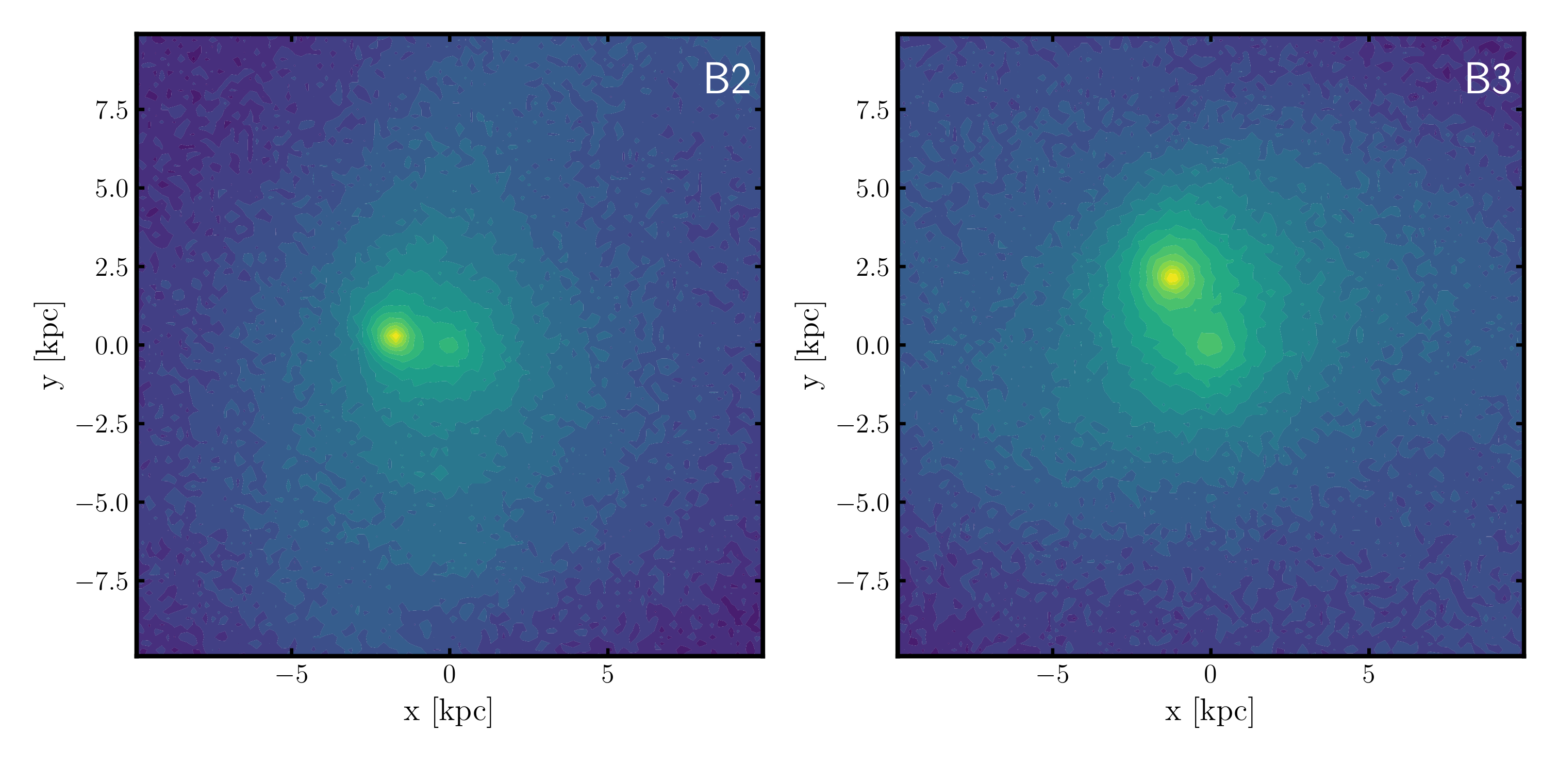}
    \includegraphics[width=2.0\columnwidth]{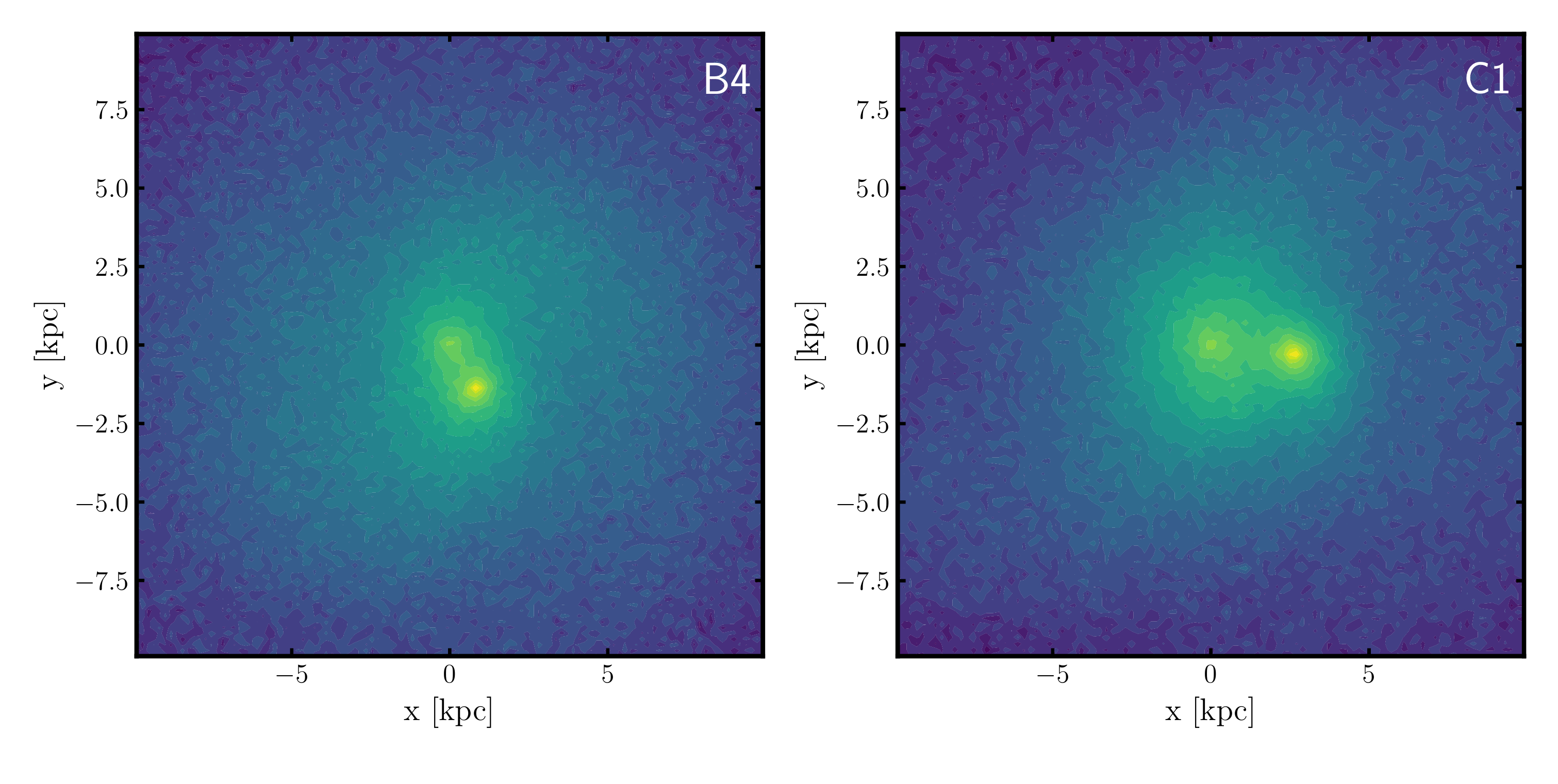}
    \caption{Stellar surface density of subsequent mergers at times where we observe bumps in the surface luminosity profiles (see Fig.\ref{fig:subsequent_bump}). The infalling satellites are represented by the brighter overdense regions. }
    \label{fig:density_contour}
\end{figure*}

The photometric analysis of A2261-BCG shows a number of objects in the core region, referred to as ``knots" \citep{postman2012,bonfini2016}. Three of these objects (knots 1-3) lie along the line of sight within the core, causing a bump in the surface brightness profile \citep[see Fig. 1 of][]{bonfini2016}. The authors suggest that, in the context of the stalled perturber scenario, the most massive knot, or all three combined, could be (at least partially) responsible for the large core observed in the galaxy as the mass contained within the core matches that of the knots. It was suggested in \cite{2017ApJ...840...31D} and \citet{Burke2017} that these knots are stalled infalling satellites which have been stripped during their infall.\newline
We plot the surface luminosity profiles in Fig.~\ref{fig:subsequent_bump} at the times where these bumps are visible. 
While the bump in model B2 is quite small, due to the smaller mass of the secondary, we see prominent bumps for the larger mass ratio simulations (B3, B4 and C1). Interestingly, they are found at radii comparable to the largest known break radii for massive ellipticals. We find that these bumps persist, on average, for about 10 Myr (see Table \ref{tab:break_radius}), so they are generally short-lived compared to the duration of the merger. However, if a galaxy is observed while a smaller secondary is infalling, such bumps can be seen in the surface luminosity profile and can make the core appear larger. 
We then plot the stellar surface density of the subsequent mergers in Fig.~\ref{fig:density_contour} at the times where we observe the bumps in the luminosity profiles. The brighter regions represent the overdense infalling secondary while the centre of the primary galaxy remains approximately at the origin. These overdense structures are seen in all simulations featuring bumps in the surface luminosity profile, and resemble the observed knots in A2261-BCG.\newline 
We estimated the mass enclosed within these knot-like structures in order to compare them with the estimated observed mass determined in \citet{bonfini2016}. To do this, we centred on the tidally stripped secondary galaxy \citep{power2003} and estimated its radial extent by numerically calculating its tidal radius according to 
\begin{equation}
    r_{t} = r_{\rm{orbit}} \left(\frac{M_{\rm{sat}}\left(<r_{t}\right)}{M_{\rm{gal}}\left(<r_{\rm{orbit}}\right)}\right)^{\frac{1}{3}}.
    \label{tidal_radius}
\end{equation}
We then adopt the stellar mass enclosed within the tidal radius as an estimate of the the spheroidal luminous mass of the knot structure, $M_{\rm{knot}}$. These are given in Table \ref{tab:break_radius}. 
We find that the mass of the secondary galaxy in remnant B2 is about half that of knot 3, whereas the secondary galaxies in the more massive remnants are larger. It is therefore plausible that the knots observed by \citet{bonfini2016} are tidally stripped satellites of independent accretion events. However, additional information like the age, metallicity, and relative velocity of the knots are required to better constrain their origin.\newline
A significant implication of our results is that knot 3 should contain a central SMBH of mass $2.54\times 10^{9}\msun \leq \msbh \leq 6.35\times 10^{9}\msun$. It is unclear whether the smaller knots should contain central SMBHs, as they are significantly less massive. We also find that merger B1 (a 1:10 minor merger with parameters from abundance matching) does not show a `bump' in the surface brightness profile due to the infalling satellite being less massive and not very dense. Low mass satellites would need to be much denser than considered here to produce a visible bump in the profile.

\subsection{Core size}
\label{sec:core}

Several different approaches can be found in the literature to identify and measure core sizes, both in simulations and observational data.
A popular method to determine the core size is to fit the surface brightness profile of the galaxy with the Nuker-profile \citep{lauer1995} 
\begin{equation}
\mu(R) = \mu_{\rm{b}} 2^{(\beta-\gamma)/\alpha} \left( \frac{R}{R_{\rm{b}}} \right)^{-\gamma} \left[ 1+\left( \frac{R}{R_{\rm{b}}} \right)^\alpha  \right]^{(\gamma-\beta)/\alpha},
\end{equation}
where $R_{\rm{b}}$ is the break radius, $\mu_{\rm{b}}$ is the surface brightness at the break radius, $\gamma$ is the inner logarithmic slope of the profile, $\beta$ is the outer logarithmic slope and $\alpha$  regulates the steepness of the transition between the outer and inner profile relative to the break radius. In this approach, once a core has been identified, the break radius $R_{\rm{b}}$ is used as a measure of the core size. However, \citep{graham2003b} suggest that the model parameters become unreliable when fit to light profiles with a large radial extent and the break radius shows a dependence on the radial fitting range.\newline 
An alternative approach is to fit the 6-parameter core-S\'ersic profile \citep{graham2003b,trujillo2004} 
\begin{equation}
\mu(R) =\mu' \left[1+\left(\frac{R_{\rm b}}{R}\right)^{\alpha}\right]^{\gamma /\alpha}
\exp \left[-b\left(\frac{R^{\alpha}+R^{\alpha}_{\rm b}}{R_{\rm e}^{\alpha}}
\right)^{1/(\alpha n)}\right], 
\label{core_sersic}
 \end{equation}
with 
\begin{equation}
\mu^{\prime} = \mu_{\rm b}2^{-\gamma /\alpha} \exp 
\left[b (2^{1/\alpha } R_{\rm b}/R_{\rm e})^{1/n}\right], 
\end{equation}
where $\mu_{\rm b}$ is the intensity at the break radius $R_{\rm b}$, $R_{\rm e}$ is the effective radius, $n$ is the  S\'ersic index in the limit $R_{\rm{b}} \to 0$ (or $R \to \infty$) and $\alpha$ regulates the steepness of the transition between the outer slope $n$ and the inner slope $\gamma$ of the S\'ersic profile.
The identification of core galaxies, and the further measurement of core sizes, may differ in the two approaches.
Break radii obtained by fitting the core-S\'ersic model tend to be smaller (up to a factor $\sim2-3$) than the Nuker break radii
\citep[e.g.][]{trujillo2004,dullograham2012,dullograham2013,dullograham2014}. As a result, ``cored" galaxies identified with the Nuker method may be classified as S\'ersic spheroids with a smaller S\'ersic index $n$ and no depleted core \citep{dullograham2014}.
A non parametric model \citep{carollo1997,lauer2007} has also been proposed to estimate the core radius, referred to as the ``cusp" radius $r_{\gamma}$. This is defined as the radius where the logarithmic slope of the surface brightness profile is equal to $-1/2$:
\begin{equation}
    \frac{\rm{d}\log \mu(R)}{\rm{d}\log R}\bigg|_{R=R_{\gamma}} = -\frac{1}{2}
\label{cusp_radius}    
\end{equation}
and is applied once the presence of a core has been established. It has been shown that this ``cusp" radius is consistent with the core-S\'ersic break radius $R_{\rm{b}}$ \citep{dullograham2012,rantala2018}, though it can be smaller, especially in the case of very flat profiles \citep{GM2012}.

\begin{table*}
\begin{center}
\caption{Break radii derived from the MCMC fitting procedure for the merger remnants described in Table \ref{tab:merger_time_params}. From left to right, the columns give: the merger remnant simulation label; the core-S\'ersic break radius at the time where $a=\af$, $R_{\rm {b1}}$; the core-S\'ersic break radius at the end of the numerical integration, $R_{\rm {b2}}$; the mass deficits $M_{\rm{def,obs1}}$ and $M_{\rm{def,obs2}}$ defined by \eqref{eq:m_def_obs} at the times where $a=\ah$ and at the end of the numerical integration; the mass deficits $M_{\rm{def,sim1}}$ and $M_{\rm{def,sim2}}$ defined by \eqref{eq:mass_def_sim} and evaluated at the same times; the average lifetime of the bump formed in the surface luminosity profile due to the infall of the secondary galaxy, $t_{\rm {bump}}$; the estimated mass of the knot according to the procedure outlined in \ref{sec:stalled_sat}, $M_{\rm{knot}}$; and the outer and inner slope parameters defined in Eq.\eqref{core_sersic} obtained from fitting the Core-S\'ersic profile at the final $N$-body integration time, $n$ and $\gamma$. Mass deficits are given in units of the total BHB mass $M_{12}$.}
\label{tab:break_radius}
\begin{tabular}{c c c c c c c c c c c} 
\hline 
\hline
Merger remnant  & $R_{\rm {b1}}$ & $R_{\rm {b2}}$ & $M_{\rm{def,obs1}} $  & $M_{\rm{def,obs2}}$ & $M_{\rm{def,sim1}}$ & $M_{\rm{def,sim2}}$ & $t_{\rm {bump}}$ & $M_{\rm{knot}}$ & $n$ & $\gamma$  \\
    & ($\rm kpc$)  & ($\rm kpc$) & ($M_{12}$) & ($M_{12}$) & ($M_{12}$) & ($M_{12}$) & ($\rm{Myrs}$) & ($10^{10}\msun$) & & \\
\hline
A1 & 0.26 & 0.60 & 0.03 & 0.26 & 0.07 & 0.28 & -  & - & 4.2 & 0.05  \\
B1 & 0.52 & 0.60 & 0.20 & 0.22 & 0.03 & 0.04 & -  & - & 4.1 & 0.19  \\
B2 & 0.44 & 0.64 & 0.15 & 0.24 & 0.0006 & 0.05 & 4.7  & 2.4 & 3.8 & 0.16 \\
B3 & 0.36 & 0.71 & 0.07 & 0.23 & -0.04 & -0.006  & 28.2 & 11.8 &3.6 & 0.09 \\
B4 & 0.35 & 0.87 & 0.06 & 0.31 & -0.018  & 0.10  & 23.5 & 13.8 & 3.6 & 0.06 \\
C1 & 0.65 & 1.01 & 0.10 & 0.27 & -0.08 & -0.05  & 18.8 & 21.4 & 3.1 & 0.07 \\
D1 & 0.33 & 0.61 & 0.02 & 0.09  & 0.017 & 0.10 & - & - & 2.5 & 0.03 \\
D2 & 0.35 & 0.90 & 0.008 & 0.13 & 0.017 & 0.09 & - & - & 2.5 & 0.02 \\
\hline
\end{tabular} 
\end{center}
\end{table*}
We choose to use the Core-S\'ersic break radius as a measure of the core size for the surface luminosity profiles obtained from our simulated mergers. This allows us to make meaningful comparisons with observed cores. Following B16 and D19, we consider a radial fitting range $0.03\,\rm{kpc}<R<100\,\rm{kpc}$.\newline
We use a Markov-Chain Monte Carlo (MCMC) approach to fit the core-S\'ersic model to the surface luminosity data including uncertainties. Because the density profiles evolve with time, we measure the break radius at two different times: $R_{\rm{b1}}$ at the time corresponding to the hard-binary separation and $R_{\rm{b2}}$ at the end of the numerical integration. These are given in Table \ref{tab:break_radius}. In the case of the major mergers, we find that larger cores are produced in D1 and D2 which use progenitor parameters from D19. This is expected based on their flatter surface luminosity profiles (see Fig.~\ref{fig:initial_merger_surface_profiles}). The cores of the major mergers continue to grow after the BHB reaches the hard-binary separation due to stellar ejections, so that $R_{\rm{b2}}>R_{\rm{b1}}$. The largest break radius, which is observed for model D2, is just under $1\kpc$, much smaller than the $\sim 2-4\kpc$ three largest observed cores. \newline
The effect of the subsequent mergers depends sensitively on the mass ratio. For the smaller mass ratio mergers B1-B2, the break radii are largely unchanged from the preceding equal mass merger A1, as expected based on the similar surface luminosity profiles. This implies that the core size in these galaxies is set by the major merger and is preserved through subsequent 1:10 mergers. The core grows slightly in models B3 (1:3 merger) and B4 (1:2 merger), with a clear dependence on the mass ratio. Model C1, which is a further 1:3 major merger from remnant B4, shows only a small increase in core size. The core radii support our earlier conclusion based on the density profiles alone that cores do not necessarily grow through subsequent mergers. It appears that the size of the core is mainly set by the first equal mass merger, and further growth depends on the mass  of the infalling satellite and the flatness of the surface density profiles. This is in apparent disagreement with the results of \citet{rantala2019}, who find that core formation is a cumulative effect over subsequent mergers.\newline
The cores produced in our sequences of mergers are smaller than the largest $R_{\rm{b}} \geq 2\kpc$ observed cores. While we see core growth over subsequent major mergers, we find that the number of major mergers required to form such large cores would be at odds with expectations from $\Lambda$CDM \citep{Naab2009,NaabOstriker2017}. 
Note that we did not model subsequent 1:1 mergers as these would be unlikely from a cosmological point of view and would require unrealistically large galaxies as secondaries.

\subsection{The mass deficit}
\label{sec:mdef}
The mass displaced by the BHB during the hardening phase is generally referred to as the ``mass deficit" and has been shown to be proportional to the combined mass of the binary \citep[e.g.][]{milosavljevicmeritt2001,graham2003b,merritt2006}. We adopt both a theoretical and observational approach to compute the mass deficit in the merger simulations. The former exploits the fact that we know the initial profile of the primary galaxy so we can compute the mass deficit as the difference in stellar mass enclosed within the core radius at a given time and in the initial profile. We then have that the mass deficit at time $t$ is
\begin{equation}
\label{eq:mass_def_sim}
 M_{\rm{def}} = M_{\star}\left(r<r_{\rm{core}}\right)\bigg|_{t=0} - M_{\star}\left(r<r_{\rm{core}}\right)\bigg|_{t}
\end{equation}
where $M_{\star}$ denotes the total mass in stars and $r_{\rm core}$ the estimated core radius.\newline
However, in order to allow a comparison with observational data we also adopt a surface-density based approach in which first the luminosity deficit is computed 
as the difference between the integrated luminosity of the inward extrapolation of the S\'ersic part of the core-S\'ersic model compared to the luminosity of the core-S\'ersic model itself \citep[e.g.][]{dullograham2014,bonfini2016,dullo2019}. 
The mass deficit is then derived by applying an appropriate mass to light ratio $\Upsilon$
\begin{equation}
\label{eq:m_def_obs}
M_{\mathrm{def}} = 2\pi \Upsilon \int_0^{R_{\mathrm{core}}} \left[ \mu_{\mathrm{s}}(r) - \mu_{\mathrm{cs}}(r) \right] r\, \mathrm{d}r,
\end{equation}
where $\mu_{\rm{s}}(r)$ and $\mu_{\rm{cs}}(r)$ are the surface luminosity profiles of the extrapolated S\'ersic profile and the core-S\'ersic profile (obtained from our MCMC fit), respectively, and we set $\Upsilon=3.5M_\odot/L_\odot$ \citep{bonfini2016}. The normalisation constant of the S\'ersic profile is obtained by equating the S\'ersic and core-S\'ersic profiles at the break radius.\newline
\begin{table*}
\begin{center}
\caption{Parameters of the GW recoil simulations. The B-GW and D-GW models are produced from the merged remnants A1 and D2, respectively. The kick velocity $\rm{v_{kick}}$ is given in  units of the escape velocity $v_{\rm{esc}}$ from the centre. The break radii $R_{\rm{b,GW1}}$ and $R_{\rm{b,GW2}}$ are  evaluated at the time when the SMBH is at the first apocentre/pericentre passage with respect to the COM of the system, while $R_{\rm{b,GW3}}$ is evaluated at the end of the numerical integration. The mass deficits $M_{\rm{def,obs3}}$ and $M_{\rm{def,sim3}}$ are calculated using equation \eqref{eq:m_def_obs} and equation \eqref{eq:mass_def_sim}, respectively, both at the end of the numerical integration. $R_{\rm {b,ini}}$ is the initial break radius of the remnant prior to SMBH ejection.
The escape velocity from the centre of A1 and D2 are $v_{\rm{esc}}\approx 3840\kms$ and $v_{\rm{esc}}\approx 4200\kms$, respectively.}
\label{tab:gw_time_params}
\begin{tabular}{c c c c c c c c c } 
\hline 
Run & $\rm{v_{kick}}$ & $R_{\rm {b,GW1}}$ & $R_{\rm {b,GW2}}$ & $R_{\rm {b,GW3}}$ & $M_{\rm{def,obs3}}$ & $M_{\rm{def,sim3}}$ & $R_{\rm {b,ini}}$ &Merger remnant \\
      & ($\rm{v_{esc}}$)  & ($\rm kpc$)  & ($\rm kpc$) & ($\rm kpc$) & ($\msbh$) & ($\msbh$) & ($\rm {kpc}$) & \\
\hline
B-GW01 & 0.1 & 0.81 & 0.82  &  0.82 & 0.50 & 0.38 &  0.60 & A1   \\
B-GW03 & 0.3 & 1.23 & 1.27  &  1.32 & 0.94  & 1.26 & 0.60 & A1  \\
B-GW06 & 0.6 & 1.27 & 1.35  &  1.42 & 1.00 & 1.51 & 0.60 & A1   \\
B-GW07 & 0.7 & 1.27 & 1.37  &  1.44 & 1.02 & 1.55 & 0.60 & A1   \\
B-GW08 & 0.8 & 1.29 & 1.37  &  1.48 & 1.05 & 1.58 & 0.60 & A1   \\
B-GW09 & 0.9 & 1.32 & 1.40  &  1.53 & 1.08 & 1.60 & 0.60 & A1   \\
D-GW03 & 0.3 & 2.21 & 2.42  &  2.64 & 2.70 & 0.86 & 0.90 & D2   \\
D-GW08 & 0.8 & 2.67 & 2.78  &  2.96 & 3.30 & 1.38 & 0.90 & D2   \\
\hline
\end{tabular} 
\end{center}
\end{table*}
We compute the mass deficit according to both approaches, which we call $M_{\rm{def,sim}}$ (where `sim' refers to simulations) and $M_{\rm{def,obs}}$ (where `obs' refers to observations) adopting the primary SMBH as centre and using the break radius obtained by our core-S\'ersic fits as the core radius. We consider two separate times, the time when the hard-binary separation is reached and the end of the simulation.
The results are listed in Table \ref{tab:break_radius} as a fraction of the combined mass of the binary $M_{12}$.\newline
The mass deficits computed with the two approaches are in good agreement for the initial 1:1 major mergers, though smaller than those found by \citet{merritt2006} in which $M_{\rm{def}}\approx 0.5M_{12}$. We attribute this discrepancy to a difference in initial conditions.\newline
The subsequent mergers, however, reveal significant differences in the two computations, with the observed mass deficits being larger than the theoretical ones. This is due to the fact that the observational approach must rely on the assumption that the the primary galaxy followed a S\'ersic profile prior to the merger, given by the inward extrapolation of the outer profile. Such assumption is clearly invalid in the case of subsequent mergers, where a core has already been carved by binary hardening in the previous major merger. The theoretical approach, on the other hand, adopts a realistic initial mass profile, as such information is readily available in the simulations. We see this effect clearly in the case of remnant B1 and A1, which show very similar surface density profiles (see Fig.\ref{fig:initial_merger_surface_profiles} and Fig.\ref{fig:subsequent_merger_surface_profiles}) but distinct estimates of the mass deficit.\newline
Finally, note that the negative mass deficits listed in Table \ref{tab:break_radius} are due to the increase in the central concentration of some models, namely B3, B4 and C1 (see Fig.\ref{fig:subsequent_merger_surface_profiles}).

\section{Gravitational wave recoil}
\label{sec:GW_section}
Anisotropic emission of gravitational waves (GWs) results in a net kick velocity imparted to coalescing massive black holes \citep[e.g.][]{berkenstein1973}. While the kick velocity is typically small ($\lesssim 200\kms$) for non-spinning BHs, it can reach $2000-5000\kms$ for large spins with particular orientations \citep{Gonzalez2007,Campanelli2007,lousto2011,Healy2018}. This implies that massive black holes can recoil at a significant fraction of the galaxy's escape velocity, or even be ejected, dragging a small cluster of stars with them \citep{merritt2009}. This stellar ``cloak" would most likely resemble a globular cluster or a sufficiently compact dwarf galaxy. The motion of a recoiling SMBH through the galaxy removes stars from the centre, turning a cuspy density profile into a core \citep[e.g.][]{meritt2004,boychainkolchin2004}. If a core has already been carved by binary scouring in a previous major merger, the motion of the SMBH enlarges the core and mass deficits as large as five times the SMBH mass can be produced \citep{gualandrismerritt2008}. This is due to the fact that the BH transfers energy to the core at each passage, and its oscillatory motion dampens much less efficiently than expected from Chandrasekhar's theory \citep{chandrasekhar1943} due to the flatness of the central profile \citep[e.g.][]{read2006}. Long lived oscillations and enlarged cores are expected in most cases, since kicks exceeding the galaxy's escape velocity are very rare \citep{gerosa2015}.\newline
While there is no confirmed detection of a recoiling SMBH yet, several candidates have been suggested over the years \citep[e.g.][]{Caldwell2014,Kalfountzou2017,chiaberge2018}. Interestingly, the galaxies with the two largest cores, IC 1101 and A2261-BCG, show an offset between the outer isophotes and the cored region, consistent with simulations of repeated passages of an SMBH through the core \citep{Dullo2017,dullo2019}. \citet{postman2012} favour the ejected SMBH scenario for the formation of the core observed from A2261-BCG and suggest that one of the knots may be a stellar cluster surrounding an ejected SMBH.\newline
In this section we study the motion of recoiling SMBHs in two of our remnants, to investigate the effects of displaced black holes on the core size.

\subsection{Numerical simulations}
We consider merger remnant A1 and D2 and combine the two black holes into one as described in section \ref{sec:multiple_mergers}. We impart a recoil kick velocity to the newly formed SMBH following the prescription in \citet{gualandrismerritt2008}, measured in units of the escape velocity $v_{\rm esc} (r) = \sqrt{-2\Phi(r)}$, where
$\Phi(r)$ is the total gravitational potential due to both the stellar and dark matter particles. The kick is given, arbitrarily, in the $x$-direction, and a correction is made for the momentum introduced into the system. The magnitude of the kick is parameterised in units of the escape velocity from the centre of the galaxy, as detailed in Table \ref{tab:gw_time_params}, as we make no assumption on the magnitude and orientation of the spins prior to coalescence. The escape velocities of the SMBHs from the centre of remnant A1 and D2 are $v_{\rm{esc}}\approx 3840 \kms$ and $v_{\rm{esc}}\approx 4200\kms$, respectively. No spin is given to the SMBH given that spin effects are only important in the immediate vicinity of the hole, a region we do not resolve.
We follow the evolution of the kicked black holes with \textsc{griffin} until either the SMBH settles at the centre of mass of the system or the integration time exceeds 2\,Gyr. The parameters of the simulations are listed in Table \ref{tab:gw_time_params}.

\subsection{Core size and mass deficit}
\begin{figure*}
    \centering
    \includegraphics[width=2.2\columnwidth]{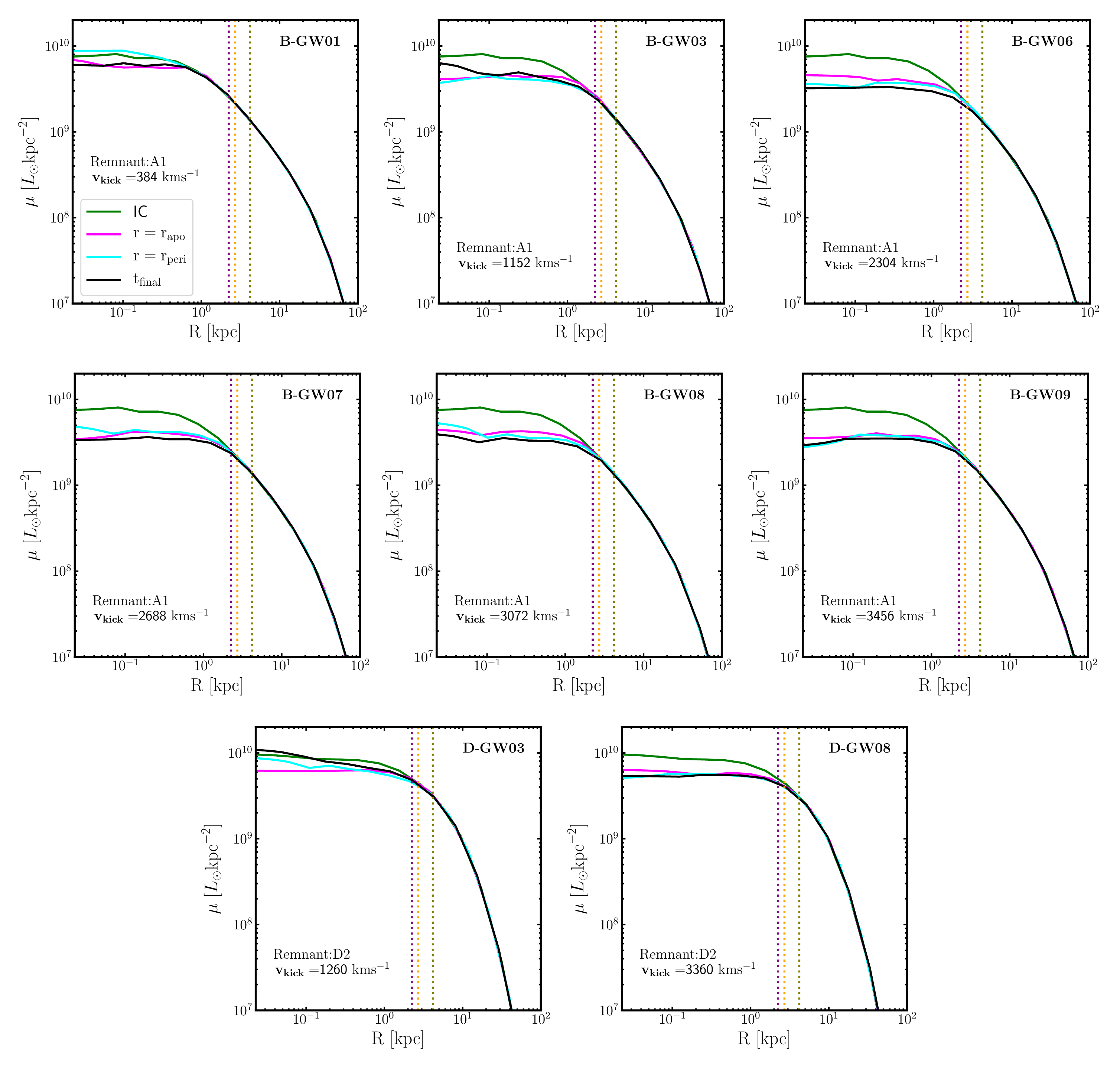}
    \caption{The surface luminosity profiles for the GW recoil simulations, with models as described in Table \ref{tab:gw_time_params}. The profiles refer to different times during the evolution: initial profile (green), the time of the maximum displacement of the SMBH from the centre of mass of the system (magenta), the time of the first return of the SMBH to the core (estimated as the time of the first pericentre with respect to the centre of mass) (cyan) and the final time of the $N$-body integration (black). The vertical dotted lines are as defined in Fig.\ref{fig:initial_merger_surface_profiles}.
   GW recoil of the SMBH causes a significant central depletion in the surface brightness profiles and enlarges the pre-existing core. In the case of the models derived using the D19 parameters, the core size obtained is comparable to the largest depleted cores known to date.
   The surface mass densities can be obtained by multiplying the surface luminosity by $\Upsilon=3.5M_\odot/L_\odot$.
   }
    \label{fig:gw_kick_surface_profiles}
\end{figure*}
Due to the combined effects of the kick and dynamical friction, the SMBHs exhibit damped oscillatory motion, in agreement with the results of \citet{gualandrismerritt2008}. For kicks larger than about $0.3 v_{\rm esc}$, the black holes reach beyond the core and experience multiple oscillations that displace stellar mass, enlarging the pre-existing core at each passage. The SMBH carries any mass bound to it in its excursions out of the core.\newline 
The surface luminosity profiles of the GW recoil models are shown in Fig.~\ref{fig:gw_kick_surface_profiles}, computed relative to the centre of mass (COM) of the system (all stars and DM particles)\footnote{We also considered the profiles centred on the SMBH and the centre of density (COD) of the stellar component. We find a very good agreement between the profiles centred on the COM and COD at all times, and also a good agreement for all three centering choice when the SMBH is at its pericentre passage within the core. We find that when the SMBH is exterior to the core region, centring on the SMBH yields nonphysical profiles.}. In order to study the evolution of the core region over time, we show the surface luminosity profiles at four distinct times: the initial profile (green); the time of maximum displacement of the SMBH from the COM (magenta); the time of the first return to the centre (cyan); and the end of the $N$-body integration (black). All models show a flattening of the inner profile and the formation of a core, roughly proportional to the magnitude of the kick. The first excursion of the SMBH to apocentre has the largest effect on the central density, followed by smaller changes with subsequent passages through the core. We observe long-lived oscillations for all cases of large kicks ($v_{\rm{kick}} = 0.6-0.9\,v_{\rm{esc}}$), in agreement with \citet{gualandrismerritt2008}, resulting in a slow but prolonged increase in the core size.\newline
We fit core-S\'ersic profiles to the surface luminosity profiles of the B-models (obtained from remnant A1) and the D-models (obtained from remnant D2) to obtain the break radii given in Table \ref{tab:gw_time_params}. Similarly to the surface luminosity profiles, these are evaluated at three significant times in the evolution: the time of the first apocentre passage of the SMBH, the time of the first return to the centre and the end of the integration. We again see evidence that most of the core is carved during the first excursion, for kicks large enough to drive the SMBH out of the core with slowly damped oscillations. For the B-GW models with kick velocities large enough to drive the SMBH out of the core, we find an increase in the final break radius of a factor $2-2.5$ with respect to the initial one, depending on the magnitude of the kick. While this represents a significant effect, it is not sufficient to explain the core size measured in A2261-BCG. For the D-GW models, on the other hand, we find break radii as large as $3\kpc$, making this mechanism a plausible explanation for the largest cores known to date.\newline
\begin{figure}
    \centering
    \includegraphics[width=1.0\columnwidth]{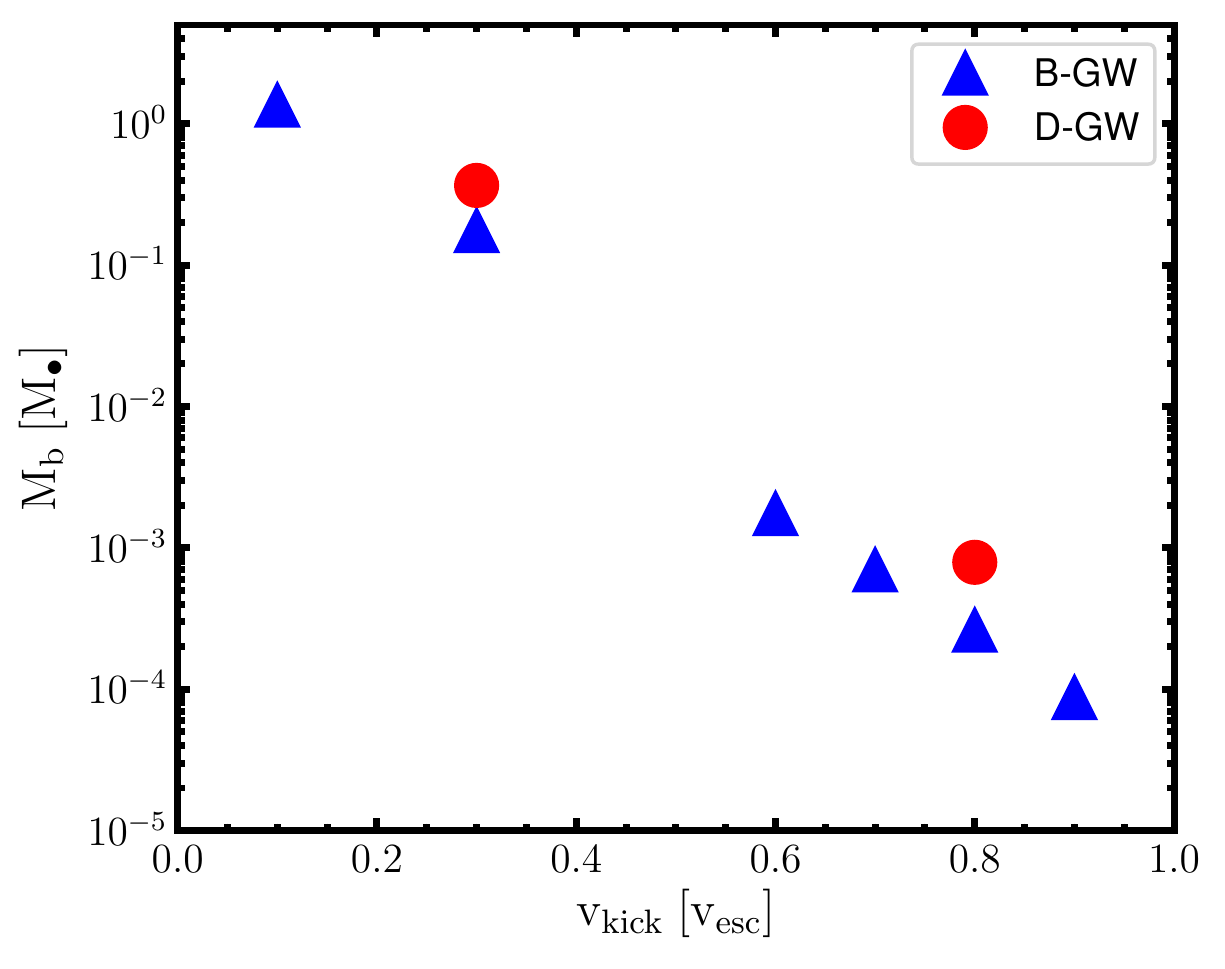}
    \caption{Initial stellar mass bound to the recoiling SMBH for the B-GW models (blue triangles) and D-GW models (red circles).}
    \label{fig:bound_mass_ic}
\end{figure}
The mass initially bound to the black hole, i.e. just after the kick is imparted, is shown in Fig.~\ref{fig:bound_mass_ic} for all models considered in Table \ref{tab:gw_time_params}. In agreement with previous studies \citep{meritt2004,boychainkolchin2004,gualandrismerritt2008, merritt2009}, it decreases steeply with increasing $\rm{v_{kick}}$. The models derived from the D2 remnant are characterised by a larger bound mass, as expected from the larger SMBH mass. However, the bound mass for such models is also larger when given in units of the SMBH mass, as shown in the figure, implying that a larger core and flatter profile also result in a larger bound mass.\newline
Interestingly, we find that most of the central surface luminosity depletion occurs by the time of the first apocentre passage the SMBH makes with respect to the cored region, where subsequent pericentre passages only cause smaller additional depletions. This behaviour is not in agreement with the results of \citet{gualandrismerritt2008}, who found that the primary mechanism of increasing the core size is the subsequent passages the SMBH makes through the core. We attribute this difference to the very different initial conditions adopted in this study, where we have multi-component models with the inclusion of a dark matter halo as well as a very large core already in place prior to the GW recoil. More mass can be bound to the SMBH in a cored profile due to the larger tidal radius, and dynamical friction is suppressed in flat density profiles. This is clearly observed in the D-GW models which start out with a flatter profile compared to the B-GW models and show a reduced effect of subsequent SMBH passages through the cores. We also find that the bound stellar mass fraction, measured when the SMBH is at apocentre, increases slowly over time, due to increasing captures as the SMBH slows down.\newline
The final core size is determined by the velocity of the kick, the SMBH mass and the slope of the central profile at the time of the recoil. We fit a power-law relation of the type
\begin{equation}
    \frac{R_{\rm{b}}}{R_{\rm{b,in}}} = K\left( \frac{\rm{v_{kick}}}{\rm{v_{esc}}} \right)^{\alpha} +1 ,
    \label{eq:break_rad_gw_kick_relation}
\end{equation}
valid for $0\leq \rm{v_{kick}} \leq \rm{v_{esc}}$,
where $R_{\rm{b,in}}$ is the break radius prior to the kick and $K$ is a constant which incorporates the dependence on both the SMBH mass and the central slope. We find best fit values of $\alpha=0.42$, $K=2.02$ in the B-GW models and $\alpha=0.18$, $K=2.41$ in the D-GW models, due to the larger SMBH mass and flatter profiles in the D2 remnant.\newline
We proceed to calculate the mass deficits at the end of the numerical integration according to Eq. \eqref{eq:mass_def_sim} and \eqref{eq:m_def_obs}. We find that these increase with increasing kick velocity (see Table \ref{tab:gw_time_params}), in agreement with previous studies \citep{meritt2004,gualandrismerritt2008,merritt2009}.
For the B-GW models and for any kick larger than $0.3\,v_{\rm esc}$, we find that the mass deficits derived from the core-S\'ersic fits are systematically smaller than the mass deficits derived from the enclosed mass. The opposite is seen in the D-GW models, where the observational approach gives mass deficits more than twice as large as the enclosed mass approach. While this is a natural consequence of the large pre-existing core present in these models, it implies that mass deficits estimated according to Eq. \eqref{eq:m_def_obs} are not a reliable measure of the mass displaced in the last merger experienced by the galaxy. Computing the mass deficit with respect to the inward extrapolation of the outer S\'ersic profiles simply provides an estimate of the mass displaced with respect to such profile, which is not a realistic assumption in the case of pre-existing cores and/or multiple mergers. We argue therefore that mass deficits computed in this way should be treated with caution when attempting to reconstruct the merger history of a galaxy, and that break radii represent a more robust measurement.\newline
We note that the cores produced by GW recoil are different, in nature, from those produced by galaxy mergers. They correspond to genuinely flat profiles in spatial density, and this is confirmed by an Abel de-projection analysis of the surface density profiles.
    
\section{Discussion}
\label{sec:discussion}

Massive galaxies have long been theorised to form via hierarchical mergers of smaller systems \citep[e.g.][]{toomre1977,white1978}. BCGs, which are found at the centre of massive galaxy clusters, are expected to have experienced significant accretion and merger events with smaller low luminosity galaxies \citep{vonderlinden2007,oliva2015}. This implies that a significant fraction of large cored galaxies are BCGs, as recently confirmed by \citet{dullo2019} who found that $\sim 77\%$ of large cored galaxies in their sample are BCGs. While cores larger than $0.5\kpc$ are rare, A2261-BCG stands out for its $2.7 - 3.6\kpc$ core \citep{bonfini2016, dullo2019}.\newline
We have explored three scenarios of core formation in massive galaxies, with the goal of explaining the very largest observed cores: binary scouring, tidal deposition and GW induced recoil. All progenitor galaxies in this study were modelled as multi-component systems including a central SMBH, a spheroidal bulge following a S\'ersic profile and a dark matter halo described by the NFW profile, with parameters selected to mimic those of A2261-BCG. All of the mergers, including both major and minor mergers, and recoil simulations were performed self consistently using the \textsc{griffin} code.

\subsection{Cores from binary scouring}

We find that binary scouring is effective in the first major merger in ejecting stellar mass from the central region, thereby lowering the central density and producing what, in projection, appears like a core. This process is responsible for turning a cuspy profile into a core, with typical sizes up to $\sim 0.9\kpc$. While binary scouring can operate in subsequent mergers, we find that its effectiveness is greatly reduced once a core is already present in the primary galaxy. Cores may then grow in subsequent mergers due to tidal deposition by massive infalling satellites.\newline
The observed size of the core cannot, therefore, be used to estimate the number of major mergers experienced by a galaxy,  as assumed in 
works such as \citet{merritt2006}. This also implies that binary scouring alone is not sufficient to explain the largest observed cores, and other mechanisms need to be invoked.

\subsection{Cores from tidal deposition and the origin of bright central `knots'}

Minor mergers, which are more abundant in typical merger trees, produce bumps in the surface luminosity profiles and can somewhat increase the size of the core due to tidal deposition. Indeed, infalling satellites represent a plausible explanation for the peculiar ``knots" observed in A2261-BCG \citep{postman2012,bonfini2016,2017ApJ...840...31D}. Bumps typically persist for a few tens of Myr in our simulations, but this will vary with the mass of the satellite.\newline
According to our simulations, the most massive knot structure (knot 3) could contain a central SMBH with an estimated mass $2.54\times 10^{9}\msun \leq \msbh \leq 6.35\times 10^{9}\msun$. Smaller knots may not contain a SMBH due to their lower mass. We find that lower mass satellites do not produce observable bumps in our simulations, and we expect that they would need to be much denser to have an observable effect.

\subsection{Cores from gravitational wave recoil and an alternative model for bright `knots'}

Recoil due to anisotropic GW emission significantly enlarges the core scoured during the phase of binary hardening. The core size scales with the kick velocity to the power $\sim 0.15-0.3$, but is also a function of the black hole mass and the slope of the central density profile, with larger cores produced in flatter profiles. Our galaxy models, adopting the structural parameters and SMBH mass by \citet{dullo2019}, naturally produce cores as large as $\sim 3\kpc$ for the largest kick velocities. We are therefore able to explain the largest observed cores if a major merger leading to binary scouring and coalescence of the holes is followed by GW recoil of the newly formed SMBH, with a kick velocity in excess of $\sim 1000\kms$. Such large kick velocities are likely very rare as they arise only from peculiar orientations of the spins of the merging black holes \citep{Gonzalez2007,Healy2018,lousto2019}. However, only four galaxies are known with core radii larger than $2\kpc$, implying that extremely large cores are also rare.\newline
Interestingly, the galaxies with the three largest cores known to date show an offset in their outer isophotes with respect to the cored region \citep{postman2012,dullo2019,gultekin2020}. An offset is naturally expected as a result of the oscillatory motion of a recoiling SMBH.\newline
The knots observed in A2261-BCG could be due to the ejection of SMBHs carrying along an envelope of bound stars, called ``Hyper-Compact Stellar Systems" (HCSS) \citep{komossa2012}. These HCSSs have sizes and luminosities similar to globular clusters (or ultra compact dwarf galaxies in extreme cases of near escape velocity kicks) but have larger velocity dispersions \citep{merritt2009}. 
They can be distinguished from tidally stripped galactic nuclei because of their compact nature and large velocity dispersions.
Interestingly, the GW kick velocity can be estimated from the observed broadening of the absorption line spectrum of a HCSS due to internal motion \citep{merritt2009}. We note that a compact cluster of stars surrounding the SMBH is not present in our initial conditions, but is a result of the oscillatory motion of the black hole through the galaxy.
The predicted luminosities of HCSSs are however much lower than those of the knots in A2261-BGC \citep{merritt2009}. The luminosity of the stellar cluster bound to the SMBH in the D-GW08 model is $\sim 1.8\times 10^{8}L_{\odot}$, a few times larger than the most luminous cases in \citep{merritt2009}. Therefore we consider this interpretation of the origin of the observed knots unlikely.
\citet{Burke2017} took HST spectra of three of the four knots to test whether a recoiling SMBH is a viable explanation.
The kinematics, colors and masses of knots 2 and 3 are consistent with infalling low-mass galaxies or tidally stripped larger galaxies.
An HCSS origin cannot be ruled out for knot 1 due to large errors in the measurement of the velocity dispersion. VLA radio imaging shows a compact off-centre radio source that appears to be the relic of an old AGN. \newline
Our largest core is somewhat smaller than that derived by \citet{bonfini2016}. However, the authors adopt a single component galaxy model for their fits, biasing their break and half-light radii towards larger values. A smaller break radius is obtained by fitting the galaxy with a multi-component core-S\'ersic spheroid and an exponential outer halo profile \citep{dullo2019}.

\subsection{The trouble with observed `mass deficits' and `cored' galaxies}

The mass deficit in massive elliptical galaxies has been used as a measure of the core size, in addition or instead of the break radius obtained from fitting a core-S\'ersic profile. While this can be computed trivially in $N$-body simulations as the difference in mass between the initial and final density profile, the observational approach must make an assumption about the initial profile. This is usually taken to be the integrated mass between inward extrapolation of the outer S\'ersic profile assuming a mass to luminosity ratio \citep[e.g.][]{dullograham2013,dullograham2014,bonfini2016}. While this is a reasonable assumption in major mergers where a core is first formed through binary scouring from a cuspy profile, it necessarily fails in subsequent mergers, both minor and major, where little mass is displaced from the central region. We therefore argue that the mass deficit cannot be directly used to constrain the merger history of a galaxy.\newline
We emphasise once more that the presence of a flat core in the surface density profile of a galaxy, which is then traditionally defined a {\it core galaxy} \citep{ferrarese1994,lauer1995}, does not imply a constant density region in the central space density. In fact, any spatial density as steep as $\rho(r) \sim r^{-1}$ projects onto a seemingly flat profile in surface density \citep{milos2002}. We find that our models are well represented by a central power-law profile of the type $\rho(r)\sim r^{-0.5}$,  in full agreement with both theoretical models \citep{mf1995} and observations \citep[e.g.][]{lauer1995}.

\subsection{The need for over-massive black holes}

The sequence of mergers that resulted in the largest cores in our simulations adopted an over-massive SMBH that lies above the $\msbh-\sigma$ and $\msbh-\rm{L}$ relations for $\msbh \geq 10^{10}\msun$ \citep{thomas2016,Dullo2017,rantala2018,dullo2019}. Recently the most massive SMBH was observed in Holm 15A with $\msbh = (4.0 \pm 0.8) \times 10^{10}\msun$ which is approximately 9 times larger than expected from the $\msbh-\sigma$ relation and 4 times larger than expected from the stellar mass relation \citep{mehrgan2019}.
Overmassive black holes are not fully understood but may arise from compact `blue nugget' galaxies at high redshift ($z\geq6$), where high velocity dispersions allow SMBHs to reach larger masses \citep{king2019}. Slow spinning SMBHs are expected to be the most massive, as these are less efficient at producing feedback in the form of outflows \citep{zubovas2019}.
The need for over-massive black holes in producing the largest observed cores is motivated by the approximately linear dependence of core sizes carved by both binary scouring \citep{merritt2006} and GW recoil on SMBH mass \citep{gualandrismerritt2008}. In principle a different sequence of mergers and/or a larger GW recoil velocity might have produced a larger core than the one obtained here. However, we don't consider these effects sufficient to produce the largest cores without an over-massive SMBH.

\subsection{Cores from SMBH triples}

An additional mechanism able to produce large cores is scouring from SMBH triples \citep[e.g.][]{hoffmanloeb2007,kulkarni2012}. These form whenever a merger with a third galaxy occurs before the coalescence of a BHB. The dynamical evolution of such triples depends sensitively on the masses and configuration of the system. If the orbit of the intruder is highly inclined with respect to that of the inner binary, Kozai-Lidov oscillations lead to high eccentricity at pericentre and trigger coalescence of the SMBHs before a close three-body encounter can take place \citep{iwasawa2006,hoffmanloeb2007}. In most other cases, a strong and often resonant encounter takes place, with significant energy transfer leading to hardening of the inner binary and ejection of one SMBH. The lightest SMBH usually escapes while the binary recoils. Escape of all three SMBHs is extremely rare \citep{hoffmanloeb2007}. $N$-body simulations of SMBH triples in gas-free major mergers show enhanced core scouring with larger mass deficits than binary scouring and up to $\sim 5 \msbh$ \citep{iwasawa2008}. \citet{kulkarni2012} find that 
triple or multiple SMBHs are more likely at high redshift and in more massive galaxies. Using $N$-body simulations with cosmologically motivated initial conditions, they show that halos with present day mass of $10^{14}(10^{15})\msun$ have a $40\% (50\%)$ probability of having more than two SMBHs at redshift $2<z<6$. Triple SMBH scouring is therefore a possible, though rare, mechanism for producing large cores in massive galaxies at high redshift.

\section{Conclusions}\label{sec:conclusion}

We have explored three scenarios of central core formation in massive galaxies, with the goal of explaining the very largest observed cores: binary scouring, tidal deposition and gravitational wave induced recoil. All progenitor galaxies in this study were modelled as multi-component systems including a central SMBH, a spheroidal bulge following a S\'ersic profile and a dark matter halo described by the NFW profile, with parameters selected to mimic those of A2261-BCG, a giant elliptical galaxy with a central surface brightness core of size $\sim 3\kpc$. All simulations were performed self-consistently using the \textsc{griffin} code.\newline
We find that we can only produce the large surface brightness core of A2261-BCG, as measured by \citet{dullo2019}, with the combination of a major merger that produces a $\sim 1\kpc$ core through binary scouring, followed by the subsequent GW recoil of the SMBH that acts to then grow the core size to $\sim 3\kpc$ as the SMBH oscillates about the galactic centre. A key prediction of this model is that the SMBH in A2261-BCG should be displaced from its centre. This would explain the offset of the outer isophotes with respect to the central region observed in the galaxies with the three largest cores \citep{postman2012,dullo2019,gultekin2020}.\newline
Our model can also explain the bright `knots' observed in the core region of A2261-BCG in one of two ways: either as the core of minor merging galaxies still on their way in or as a bound cluster of stars surrounding previously ejected SMBHs. 
The GW recoil model predicts a smaller and more tightly bound star cluster at the location of each knot than the `minor merger' model that predicts a more extended and kinematically hotter distribution of stars \citep{merritt2009}. Because
HCSSs are predicted to have much lower luminosities than the knots in A2261-BGC \citep{merritt2009}, we find the `minor merger' scenario a more likely explanation for the knots.\newline
We find that observed `mass deficit' calculations in the literature are not reliable. These are typically calculated as the difference in stellar mass between the observed central surface brightness core and an inward extrapolation of an outer S\'ersic profile fit. This is a reasonable assumption following a single major merger (in which a central core is formed through binary scouring alone). However, the calculation necessarily fails following subsequent mergers, both minor and major. Thus, the observed `mass deficit' at the centres of giant elliptical galaxies, derived in the above way, cannot be directly used to constrain their merger histories.\newline
Finally, we confirm with simulations that the cored surface brightness profiles produced by a sequence of dry galaxy mergers are in fact weak cusps in spatial density, and only appear cored in projection, in agreement with prior theoretical \citep{mf1995} and observational \citep{ferrarese1994, lauer1995, gebhardt1996} studies. Cores produced by GW recoil, on the other hand, result in genuinely flat 3D density profiles. Non-divergent central profiles are only observed for a few core galaxies \citep{lauer2002,lauer2005}, and A2261-BCG is one of them \citep{postman2012}, lending further support to our proposed scenario.

\section*{Acknowledgements}
We thank Eugene Vasiliev for insightful discussions regarding the initial conditions for the galaxy models and Fani Dosopoulou for illuminating discussions about the core properties of A2261-BCG.
MD acknowledges support by ERC-Syg 810218 WHOLE SUN.
FA acknowledges support from a Rutherford
fellowship (ST/P00492X/1) from the Science and Technology Facilities Council. 

\section*{Data availability}
The data underlying this article will be shared on reasonable request to the corresponding author.



%\bibliographystyle{mnras}

\bibliographystyle{mnras}
\bibliography{main}

% Don't change these lines
\bsp	% typesetting comment
\label{lastpage}
\end{document}